\documentclass[fleqn,usenatbib]{mnras}

\usepackage{newtxtext}%,newtxmath}
\usepackage{mathptmx}
\usepackage{txfonts}

\usepackage[T1]{fontenc}
\usepackage{ae,aecompl}
\usepackage{hyperref}

\usepackage{graphicx}	% Including figure files
\usepackage{amssymb} % Extra maths symbols
\usepackage{footnote}
\usepackage{adjustbox}
\usepackage{lipsum}
\usepackage{mwe}
\usepackage{verbatim}
\usepackage{blindtext}
\usepackage[group-separator={,},group-minimum-digits=4]{siunitx}
\usepackage[normalem]{ulem}
\usepackage{subfig}

\title[UVIT observation of NGC 7492]{Ultraviolet Imaging Telescope (UVIT) observation of the Galactic Globular Cluster NGC 7492}

\author[Kumar et al.]{
Ranjan Kumar,$^{1}$\thanks{E-mail: ranjankmr488@gmail.com}
Ananta C. Pradhan,$^{1}$\thanks{E-mail: acp.phy@gmail.com}
Abhisek Mohapatra,$^{1}$
Ayush Moharana,$^{1,2}$
\newauthor Devendra K. Ojha,$^{3}$ 
M. Parthasarathy,$^{4}$
and Jayant Murthy $^{4}$ \\
$^{1}$Department of Physics and Astronomy, National Institute of Technology, Rourkela, Odisha - 769 008, India\\
$^{2}$Nicolaus Copernicus Astronomical Center, Polish Academy of Sciences, ul. Rabia\'{n}ska 8, 87-100 Toru\'{n}, Poland \\
$^{3}$Department of Astronomy and Astrophysics, Tata Institute of Fundamental Research (TIFR), Mumbai - 400 005, India\\
$^{4}$Indian Institute of Astrophysics, Bangalore - 560 034, India\\
}

\date{Accepted 2020 December 22. Received 2020 December 22; in original form 2020 March 13}

\pubyear{2020}

\begin{document}
\label{firstpage}
\pagerange{\pageref{firstpage}--\pageref{lastpage}}
\maketitle

% Abstract of the paper
\begin{abstract}
We present detailed photometric observations of the Galactic globular cluster NGC 7492 using the data obtained with two far-ultraviolet (FUV: 1300 - 1800 \AA) and three near-ultraviolet (NUV: 2000 - 3000 \AA) filters of Ultraviolet Imaging Telescope (UVIT) on-board the \textit{AstroSat} satellite. We confirmed the cluster membership of the extracted sources using GAIA data release 2 (Gaia DR2) proper motion data. We have used color-magnitude diagrams (CMDs) using UVIT and GAIA filters to separate out different evolutionary stages of the stars present in the cluster. We have identified a new extreme horizontal branch (EHB) star at the core of the cluster using UV and UV-optical CMDs. The estimated distance-modulus of the cluster is $16.95\pm0.05$ obtained by fitting BaSTI isochrones with cluster parameters, $[Fe/H] = -1.8$ dex and age $= 12.0$ Gyr on the V $-$ I vs V CMD. Interestingly, only the EHB star and blue horizontal branch stars (BHBs) among the UV-bright hot sources are detected in FUV filters of UVIT. We have derived the effective temperature of BHBs using color-temperature relation and spectral energy distributions (SEDs) of multi-band filters, which are in the range from 8,000 K to 10,500 K. We find a variation of He abundance of BHBs by fitting the BaSTI ZAHB. The range in the He abundance of the BHBs corresponding to the best fit isochrones is 0.247 to 0.350. We have estimated various physical parameters of the newly identified EHB star in the cluster using SED fit and post-HB evolutionary tracks. We have studied  the radial distribution of all the sources of the cluster detected in UVIT. The sources detected in FUV filters extend beyond the half light radius (1.15$'$) of the cluster, whereas the sources detected in NUV filters extend beyond the tidal radius (9.2$'$) of the cluster.
\end{abstract}


%%%%%%%%%%%%%%%%%%%%%%%%%%%%%%%%%%%%%%%%%%%%%%%%%%

%%%%%%%%%%%%%%%%% BODY OF PAPER %%%%%%%%%%%%%%%%%%

\section{Introduction}
\label{intro}
Galactic globular clusters (GGCs) are comprised of old stellar populations which are crucial to understand the stellar evolution as well as numerous peculiar properties of the Milky Way Galaxy. GGCs also help in constraining the structural, kinematical and dynamical properties of the Galactic halo \citep{Harris1979, Freeman1981, Rosenberg2000, Keller2012, Arakelyan2018, Posti2019, Kruijssen2019, Massari2019}. They are relatively proximate objects consisting of a huge number of coeval stars with less interstellar extinction which allows to explore the various evolutionary phases of individual stars. Subsequently, a detailed study of different stellar populations provides clues about the formation and evolution of GGCs \citep{Ferraro2003, Tenorio2016, Tenorio2019}. 

The ultraviolet (UV) light in GGCs is due to the hot, bright sources such as post-asymptotic giant branch stars (pAGB), blue straggler stars (BSs), extreme blue and blue horizontal branch stars (EHBs, BHBs), white dwarfs (WDs), etc. Many studies have performed photometric and spectroscopic analysis of these sources, which  dominate the total integrated light of the GGCs in the UV  \citep{Zinn1972, Harris1983, Moehler1998, Ambika2004, Jasniewicz2004, Moehler2010, Dalessandro2011, Schiavon2012, Dalessandro2013, Piotto2015, Subramaniam2017, sahu288, Moehler2019}. The UV bright stars in globular clusters also serve as standard candles because they have same absolute luminosity over a wide range of effective temperature (T$_{\mathrm{eff}}$) values \citep{Bond2001, Jasniewicz2009a, Parthasarathy2012}. A comprehensive and homogeneous database of magnitudes and colors of the UV bright sources for a good number of GGCs has been cataloged using observations from \textit{Galaxy Evolution Explorer (GALEX)} \citep{Schiavon2012, Dalessandro2012}, and the {\em Hubble Space Telescope (HST)} \citep{Piotto2015, hb_temp_Lagioia_2015}. Ultraviolet Imaging Telescope (UVIT) with unprecedented resolution is providing a very good opportunity to collect UV data of GGCs to unravel the complexities involved in the UV properties of the stars at the end stage of their evolution.  Recently, the UV properties of a list of GGCs have been explored using UVIT observations \citep{Subramaniam2017, sahu288, Jain2019,Kumar2020a,Kumar2020}. The authors have identified multiple stellar populations in GGCs and subsequently studied their distributions and properties using multi-band filters of UVIT. 

GGCs have been classified from class I - XII based on the concentration of sources towards their centers \citep{Shapley1918, Shapley1927}. NGC 7492 is a sparse Galactic halo globular cluster, classified as class `XII', as the distribution of the sources in the cluster are not strongly concentrated in the center of the cluster. It is situated at a heliocentric distance of about $26.3 \pm 2.3$ kpc \citep{Cote1991, Figuera2013, Carballo2018} with Galactic coordinates, $l = 53.39^\circ$ and $b = -63.48^\circ$. The age and metallicity ($[Fe/H]$) of the cluster are about $12\pm0.5$ Gyr and $-1.8\pm0.3$ dex, respectively \citep{Cohen2005, Forbes2010, Harris2010, Figuera2013}. The latest 3D extinction map of \citet{Green2019} gives the amount of visual extinction in the cluster direction at a distance of 25 kpc to be $E(B-V)= 0.04$ mag. The overall shape of NGC 7492 is flattened with a tail like structure around it as predicted by \citet{Lee2004} and latter confirmed by \citet{Navarrete2017} using the data of Pan-STARRS survey. Since, the cluster lies together with  the stellar streams of the Sagittarius dwarf galaxy, it was predicted that the origin of the cluster might be  from the Sagittarius dwarf galaxy but further investigations ruled out this claim \citep{Carballo2014, Carballo2018}.  

The UV bright sources in NGC 7492 have several studies in optical bands \citep{Buonanno1987, Cote1991, Lee2004, Carballo2012, Figuera2013} prior to their analysis in UV using {\em GALEX} observation \citep{Schiavon2012, Dalessandro2012}. A  uniform distribution of BHBs  was suggested by \cite{Buonanno1987} at the core of the cluster without any extended blue tails (i.e., no EHBs). \citet{Navarrete2017} with a Pan-STARRS wide field imaging survey found the presence of BHBs in the extended tidal arms of the cluster. \citet{Cote1991} identified 27 BSs within the cluster which were predicted to be formed due to mass segregation in the cluster. So far, seven variable sources (three RR Lyraes, two SX Phoenicis (SXPhes) and two long period variables) have been discovered in the cluster in optical bands \citep{Shapley1920, Barnes1968, Figuera2013}.

There are only a few studies of the cluster in UV bands, all using the {\em GALEX} observations \citep{Schiavon2012, Dalessandro2012}. They have provided UV CMDs for the detected UV-bright sources and integrated magnitude of the cluster. Their study was focused far off the central region of the cluster owing to the crowding effect and also they did not compare the observations with stellar evolutionary models. UVIT with superior resolution ($\sim$1.5$''$) than {\em GALEX} ($\sim$4.5$''$), has the ability to resolved the central part of the cluster establishing its better capabilities to study the GGCs. UVIT observation of the cluster has enabled us to constrain the cluster parameters by comparing stellar evolutionary models to observations.  With the intention of providing multi-band UV photometric properties of this cluster, we have observed NGC 7492 in five different filters of UVIT. In this paper, we present a detailed UV photometric analysis of NGC 7492  using two far-UV (FUV) and three near-UV (NUV) filters of UVIT on-board the Indian satellite \textit{AstroSat}.  

In Section \ref{sec:observation}, we present the observation details, data reduction process and photometry of the detected sources. In section \ref{sec:CMD}, we show the CMDs of all the detected sources. In Section \ref{sec:param}, we estimate the distance of the cluster and the He abundance of the BHBs. In Section \ref{sec:bhb}, we discuss about the estimation of T$_{\mathrm{eff}}$ of the BHBs. In Section \ref{sec:ehb}, we derive the various physical parameters of the newly identified EHB star. In Section \ref{sec:discussion}, we have discussed the spatial distribution of various sources, and then we summarize our results in Section \ref{conclusion}.

\section{Observations and Data reduction}
\label{sec:observation}

\subsection{Observations}
We have observed NGC 7492 with the UVIT instrument on-board the \textit{AstroSat} satellite. UVIT consists of 38 cm twin telescopes: one observes in the FUV (1300 - 1800 \AA) and the other in the NUV (2000 - 3000 \AA) and visible (3200 - 5500 \AA) bands using a dichroic mirror beam-splitter. The FUV and NUV channels have five filters each for imaging purpose in photon counting mode. The visible channel has five filters, observing in an integrating mode, but is primarily used for tracking purposes. The field of view (FoV) of UVIT is 30$'$ with image pixel size of 0.417$''$/pixel. The point spread functions (PSFs) of FUV and NUV filters are 1.5$''$ and 1.2$''$, respectively \citep{Tandon2017, Rahna2017}. The filter specifications of UVIT are available in UVIT - \textit{AstroSat} web-page\footnote{ \href{https://uvit.iiap.res.in/Instrument/Filters}{https://uvit.iiap.res.in/Instrument/Filters}}, whereas the instrumentation and ground based calibration are given in \cite{Kumar2012, Kumar2012a}. The on-board performance verification and in-orbits calibrations of various filters of UVIT can be found in \citet{Subramaniam2016, Tandon2017, Rahna2017, Tandon2020}.

We obtained observations with two FUV filters (BaF2 and Silica) and three NUV filters (NUVB15, NUVB13 and NUVB4) of UVIT with exposure times of 0.347 ks to 3.276 ks, depending on the filter, in the A02 cycle of observation (Observation ID: $A02\_72$, PI: Ananta C. Pradhan). The observational details are given in \autoref{tab:observation}.  

\subsection{Data Reduction}
\label{reduction}
The data reduction of UVIT observation was performed with the software package CCDLAB, written specifically for UVIT data by \citet{Postma2017}. CCDLAB converts the Level 1 data, obtained from the  Indian Space Science Data Centre (ISSDC), into astronomical images for each orbit of the five filters. We then aligned and co-added all the orbits of a filter to obtain its corresponding science image to perform the photometry. We applied astrometry from GAIA DR2 catalog \citep{GaiaCatalog2018} using  IRAF \textit{ccmap} package.

\begin{figure*}
    \centering
    \begin{minipage}{1.0\columnwidth}
        \centering
        \includegraphics[width=\columnwidth]{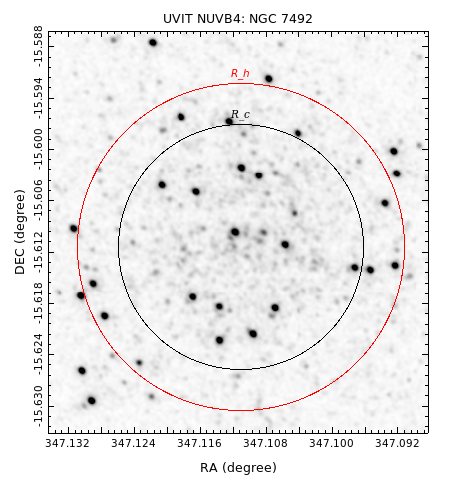} % first figure itself
    \end{minipage}
    \begin{minipage}{1.0\columnwidth}
        \centering
        \includegraphics[width=\columnwidth]{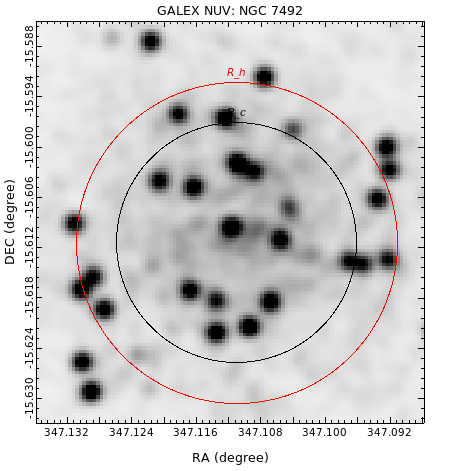} % second figure itself
    \end{minipage}
    \caption{UVIT NUVB4 ($\lambda_{\mathrm{eff}}$ = $2632$ \AA,\ \textit{left}) and {\em GALEX} NUV ($\lambda_{\mathrm{eff}}$ = $2304$ \AA,\ \textit{right}) images of NGC 7492. The red circle denotes the half light radius ($R_h$ = 1.15$'$) and the black circle denotes the core radius ($R_c$ = 0.86$'$) of the cluster \citep{Harris2010}.}
    \label{fig:im_comp}
\end{figure*}

\begin{table}
\centering
\caption{Observational details of NGC 7492.}
\label{tab:observation}
\resizebox{\columnwidth}{!}{
\begin{tabular}{lcccr}
\hline
\hline
 Date of observation & \multicolumn{4}{c}{ 2016, October 19} \\
 Telescope pointing &  \multicolumn{4}{c}{ $l=53.4^\circ$,\hspace{10pt}$b=-63.5^\circ$}\\
 \hline
Filter & mean $\lambda$ & $\Delta \lambda$ &  Exposure Time & No. of orbits \\
 & (\AA) & (\AA) &  (sec.) \\
 \hline
NUVB4 & 2632 & 275 & 3014 & 3 \\
NUVB13 &  2447 & 280 & 3276 & 3  \\
NUVB15 & 2196 & 270 & 347 & 1 \\
Silica & 1717 & 125 & 850 & 2 \\
BaF2 & 1541 & 380 & 349 & 1 \\
\hline
\end{tabular}
}
\end{table}

In \autoref{fig:im_comp}, we have compared the UVIT NUVB4 image of the crowded core region of NGC 7492 with that of {\em GALEX} NUV filter. The exposure time of UVIT NUVB4 filter is 3014 seconds and that for {\em GALEX} NUV filter is 3200 seconds \citep{Schiavon2012}. We clearly see the effect of high resolution of UVIT ($\sim$1.5$''$) over {\em GALEX} ($\sim$4.5$''$), as UV bright sources have been better resolved within the half light radius of the cluster \citep[$R_h$ = 1.15$'$;][]{Harris2010}. We see three bright sources, which are not resolved in {\em GALEX} image (right panel, \autoref{fig:im_comp}), whereas these sources are clearly separable in UVIT image (left panel, \autoref{fig:im_comp}).

\subsection{Photometry}
We have performed photometry using the IRAF \textit{DAOPHOT} package which selects a few isolated sources from the image and develops a PSF model. The model PSF is applied to all the detected sources in order to perform photometry on the entire image. We used the following steps to obtain photometry on the observed images. We generated the model PSF using $20-25$ isolated sources in each filter image. An average FWHM of 1.7$''$ and  1.5$''$ for the PSF model was obtained for the FUV and NUV filter images, respectively. We used these values to perform aperture photometry. The model PSF and aperture photometry were incorporated into the  \textit{ALLSTAR} routine to obtain the final PSF magnitudes of the detected sources. We performed a curve of growth analysis on PSF modeled sources to obtain an aperture correction value which was applied to the PSF magnitudes of the sources. Then a final catalog of apparent magnitudes for the detected sources of NGC 7492 was obtained in the FUV and NUV filters. The magnitudes of the sources in the UVIT filters were corrected for extinction values, which were estimated using \cite{cardeli1989} extinction law and A$_\mathrm{V}$ values from \cite{Green2019}.

\subsection{Other archival data}
We have used several available archival observations of the cluster for the study of CMDs and spectral energy distribution (SED) fitting of UVIT detected sources. We have obtained the proper motions ($\mu_{RA},\, \mu_{DEC}$) and G, BP and RP band magnitudes for sources using Gaia Data Release 2 \citep[Gaia DR2, ][]{GaiaCatalog2018}. Apart from the Gaia DR2 data, we have also used {\em GALEX}, Canada-France-Hawaii Telescope (CFHT), Panoramic Survey Telescope and Rapid Response System (PanSTARRs), and ground based UBVRI observations for the comprehensive analysis of the sources contained in the cluster within the UVIT FoV. In order to make a comparative study of {\em GALEX} observations with UVIT, we collected {\em GALEX} images of the cluster from the Mikulski Archive for Space Telescopes (MAST) site and the FUV and NUV magnitudes from \citet{Schiavon2012}\footnote{\href{http://www.cosmic-lab.eu/uvggc/archive.php}{http://www.cosmic-lab.eu/uvggc/archive.php}}. Similarly, we have used two filters, g and r of CFHT observation \citep{Munoz2018} and five filters  (g, r, i, y, and $z$) of PanSTARRs-PS1 survey data \citep{Chambers2016} for the SED fitting of BHBs. We have also adopted data in $U,B,V,R$ and $I$ filters from the latest released ground based archival data of the cluster \citep{Stetson2019} to study the UV-optical CMDs and SED fittings of BHBs.

\subsection{Cluster membership with Gaia DR2 proper motion}
Recent works use Gaia DR2 proper motion catalog to identify the cluster members in GGCs \citep{gaia_kinematics, Vasiliev2018, Baumgardt2019, Fierro2019}. Generally, proper motions in RA ($\mu_{RA}$) and DEC ($\mu_{DEC}$) are used to separate out the contamination of field stars from the cluster members \citep{gaia_kinematics}. The average proper motions obtained for NGC 7492 using GAIA DR2 data are ($\mu_{RA},\, \mu_{DEC}) = (0.799, -2.273$)  mas/yr \citep{Vasiliev2018, Baumgardt2019}. We cross-matched the UVIT sources with a catalog of the GAIA DR2 within a matching radius of $1.5''$. We found 424 sources in NUVB4, 322 sources in NUVB13, 67 sources in NUVB15, 42 sources in Silica and 45 sources in BaF2 filters have counterparts in the Gaia DR2 catalog. We obtained the proper motions ($\mu_{RA},\, \mu_{DEC}$) and optical magnitudes (G, BP and RP bands) from the catalog and converted the G, BP and RP band magnitudes into AB-magnitudes using zero point fluxes in AB-magnitude and Vega magnitude systems given for respective filters at Spanish Virtual Observatory (SVO) filter profile service\footnote{\href{http://svo2.cab.inta-csic.es/theory/fps/}{http://svo2.cab.inta-csic.es/theory/fps/}}. We estimated the mean $\mu_{RA}$ and mean $\mu_{DEC}$ as 0.799 mas/yr and $-$2.338 mas/yr, respectively, using Gaussian dispersion over vector-point diagram (VPD) of all the cross-matched sources. Finally, we were left with a total of 175 sources in NUVB4, 128 sources in NUVB13, 41 sources in NUVB15 and 40 sources in each Silica and BaF2 filters as cluster members.

\section{Color-Magnitude Diagrams}
\label{sec:CMD}

\begin{figure*}
    \centering
        \includegraphics[width=\columnwidth]{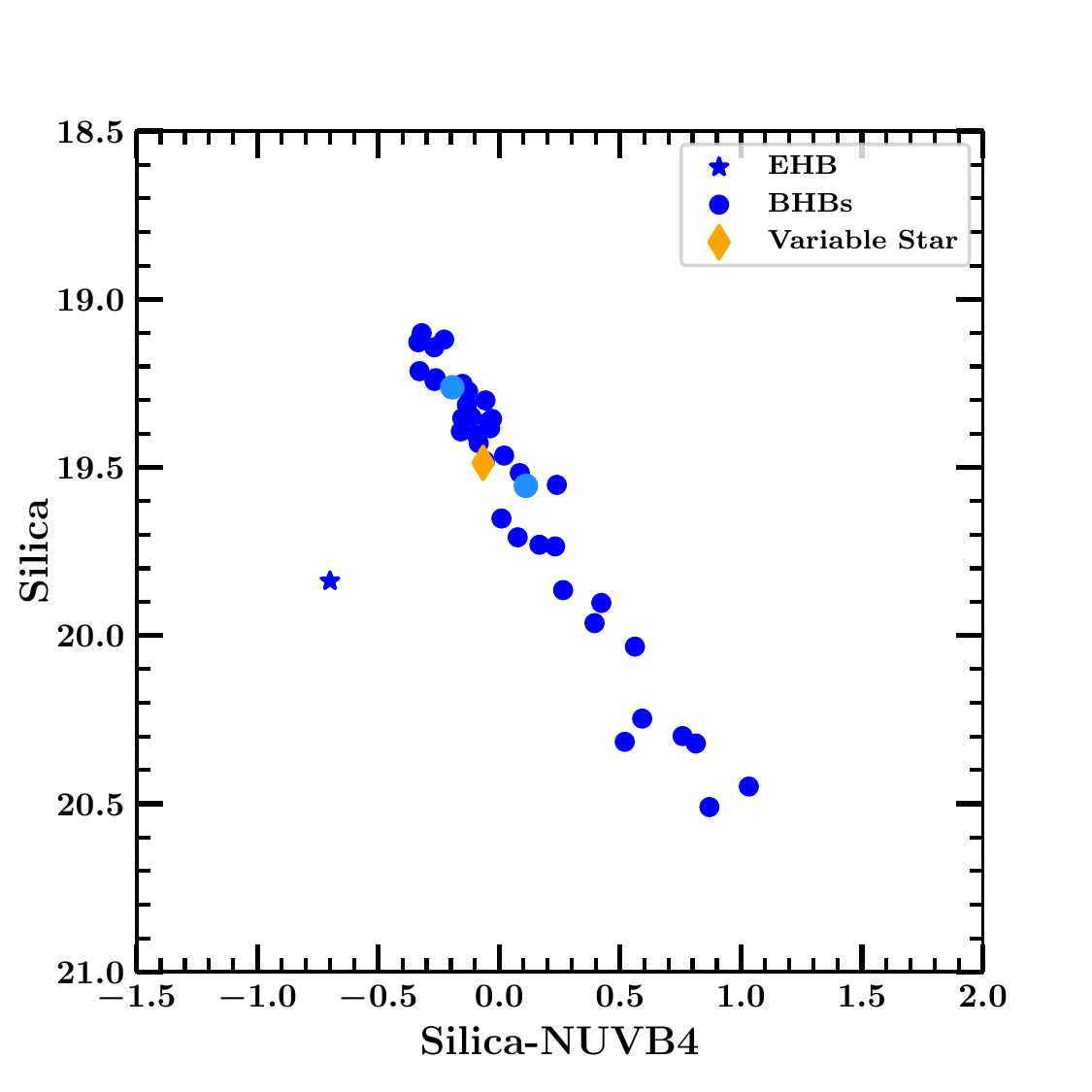}
        \includegraphics[width=\columnwidth]{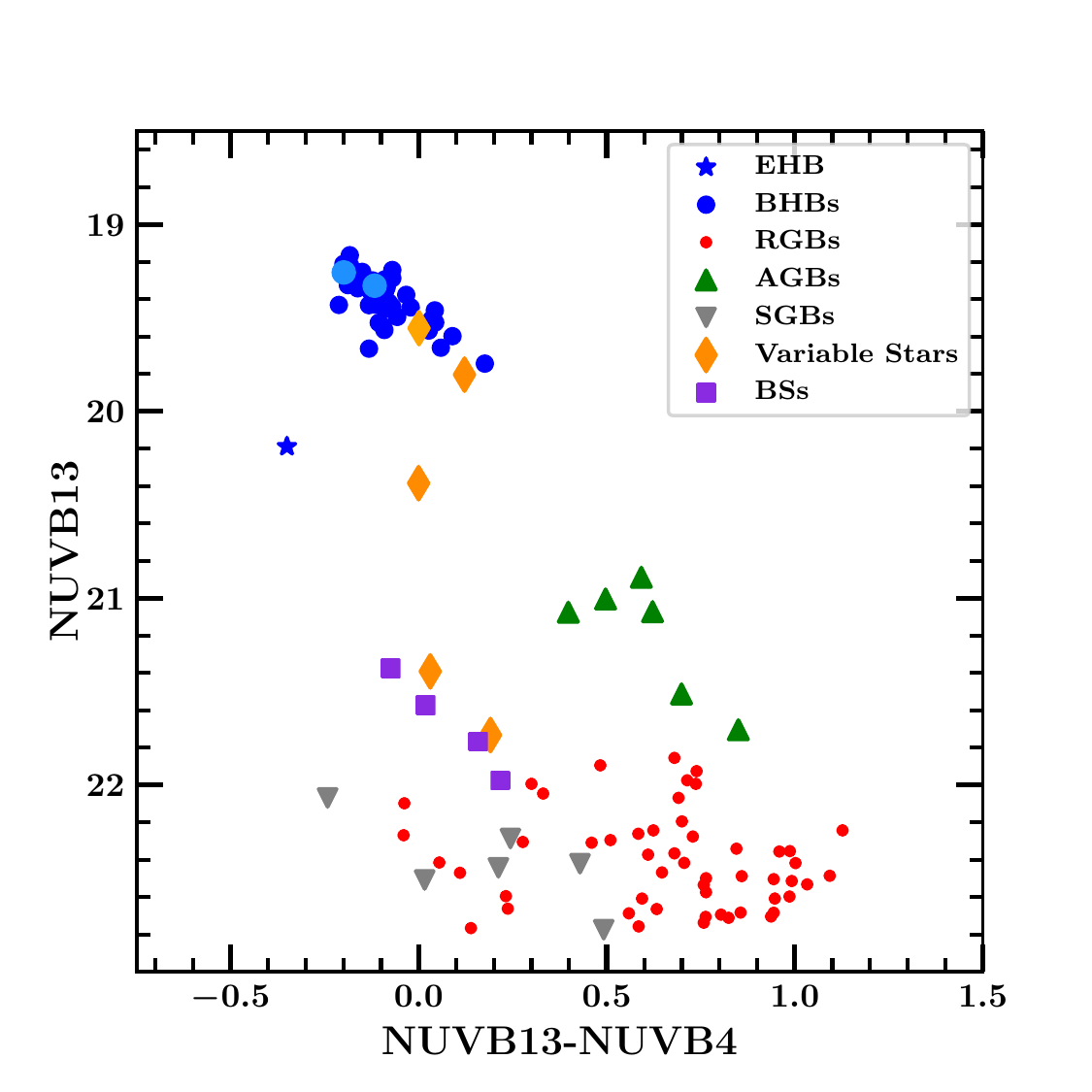}\\
        \includegraphics[width=\columnwidth]{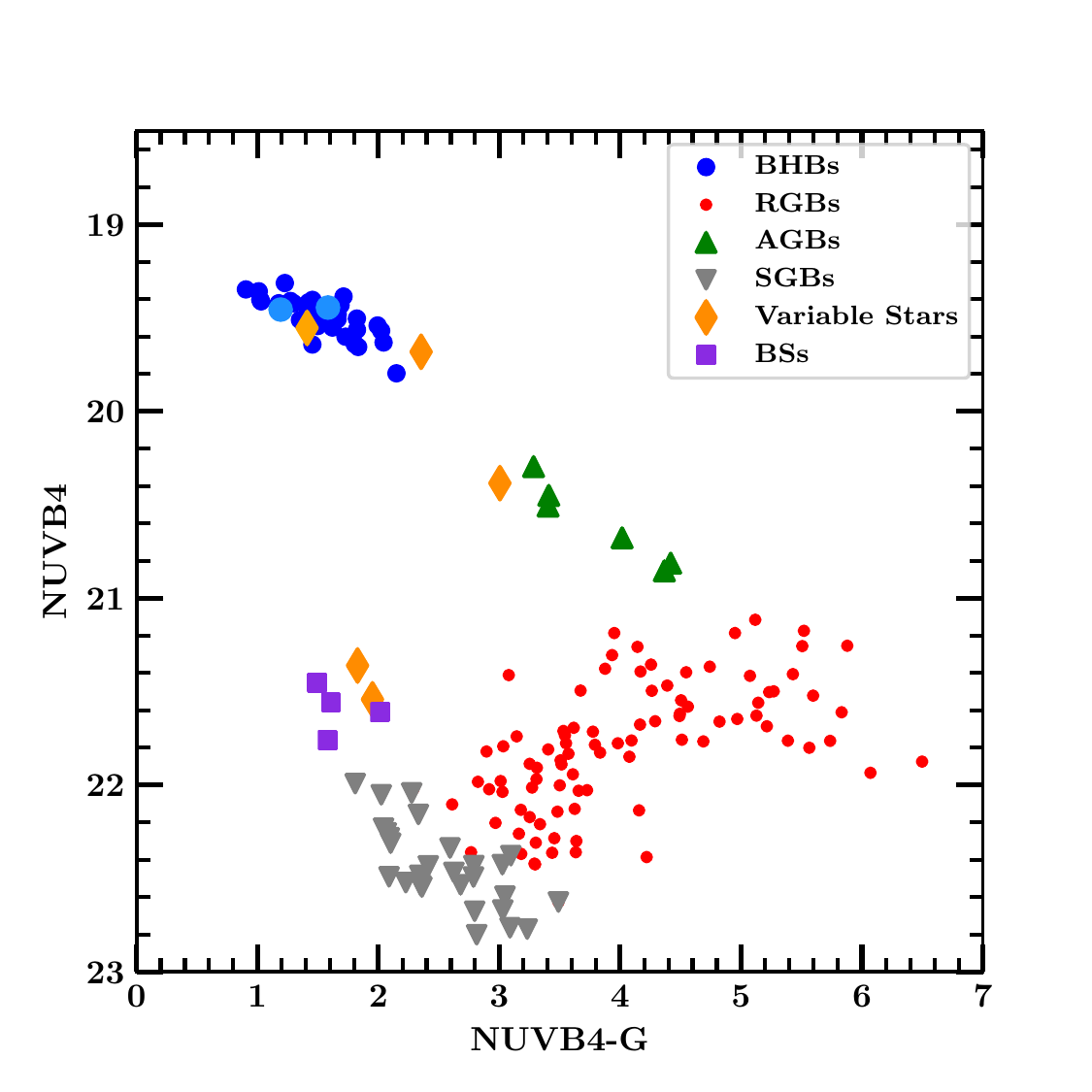} 
        \includegraphics[width=\columnwidth]{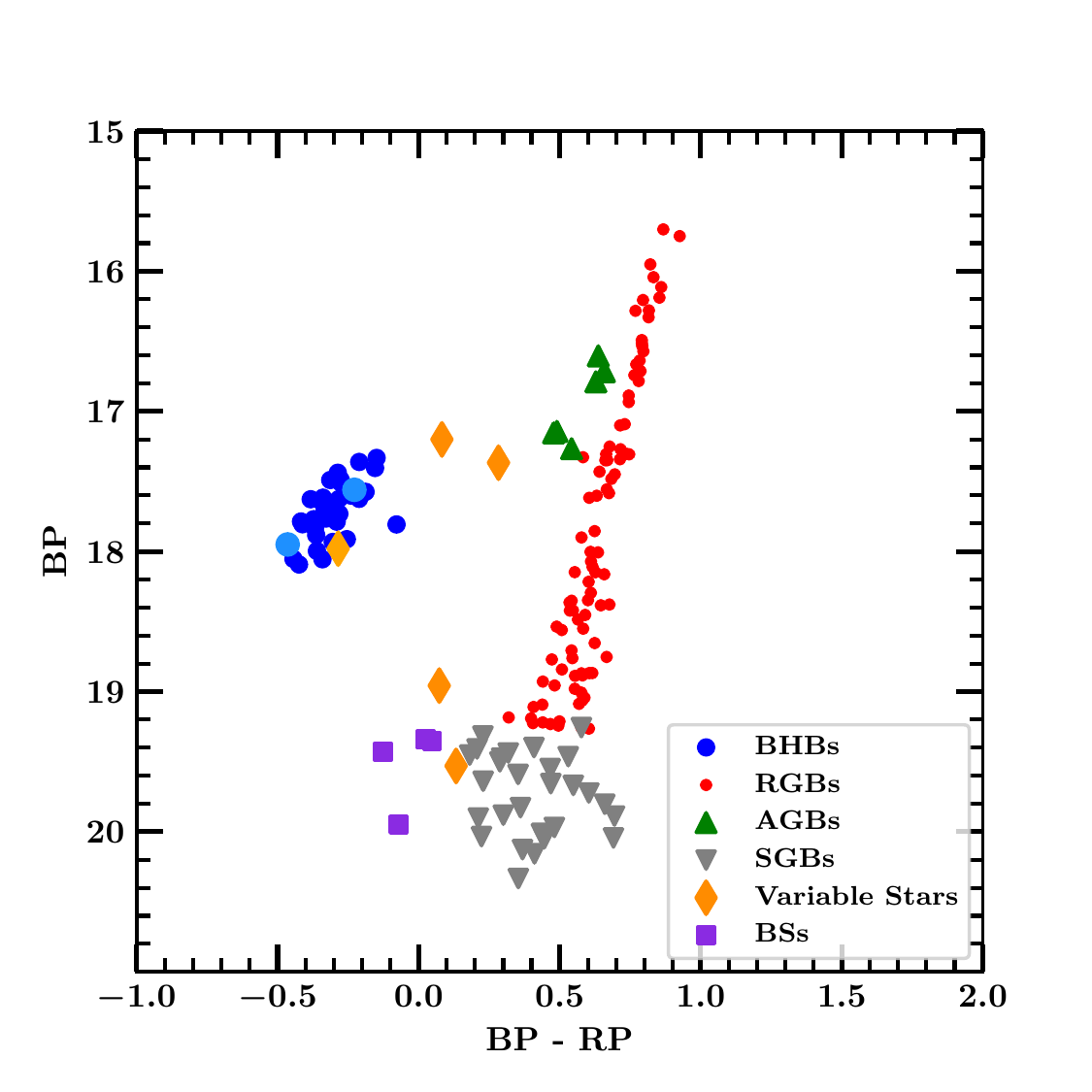}
    \caption{Silica$-$NUVB4 vs Silica CMD (upper left panel), NUVB13$-$NUVB4 vs NUVB13 CMD (upper right panel), NUVB4$-$G vs NUVB4 CMD (lower left panel) and BP$-$RP vs BP CMD (lower right panel) of the observed cluster members. EHB star is marked in blue asterisk, BHBs with Gaia proper motions are marked with blue solid points, whereas BHBs without proper motions are marked in light blue solid points. RGBs, AGBs, SGBs, variable stars and BSs are marked with red solid circles, upper solid triangles, gray lower solid triangles, and orange diamonds and violet solid squares, respectively. }
    \label{fig:uvcmd}
\end{figure*}

\cite{Schiavon2012} have found a dominant blue horizontal feature in the UV CMDs for NGC 7492 using the {\em GALEX} data. We have moved a step further in studying the details about the different evolutionary stages of the cluster members in UV and UV$-$optical CMDs. We have plotted Silica$-$NUVB4 vs Silica, NUVB13$-$NUVB4 vs NUVB13, NUVB4$-$G vs NUVB4, BP$-$RP vs BP CMDs for all the confirmed sources in \autoref{fig:uvcmd}. The stars of different evolutionary stages of the cluster were predominantly identified based upon their position in NUVB4$-$G vs NUVB4 and BP$-$RP vs BP CMDs. We found 39 BHBs, one EHB star and one variable star in FUV filters (BaF2 and Silica) and 39 BHBs, one EHB star, 4 BSs, 5 variable stars, 6 AGBs, 91 RGBs and 28 SGBs sources in NUV filters (NUVB15, NUVB13 and NUVB4). The morphology of different evolutionary stages visible in UV and UV-optical CMDs is given in the following sub-sections.

\begin{table*}
    \centering
    \caption{Details of 39 BHBs and one EHB star of NGC 7492 observed with UVIT filters. In the first column, we have defined a unique ID for each BHB and EHB. The extinction corrected magnitudes of UVIT and GAIA G filters are given here. The given T$_{\mathrm{eff}}$ with error in T$_{\mathrm{eff}}$ as $\pm$125 K of BHBs and EHB are derived from the SED fit.}
    \label{tab:BHB_catalog_apendix}
    
    \adjustbox{max width=\textwidth}{
\begin{tabular}{|l|l|c|c|c|c|c|c|c|c|c|l|c|}
\hline
  \multicolumn{1}{|c|}{ObjID} &
  \multicolumn{1}{c|}{RA(J2000)} &
  \multicolumn{1}{c|}{DEC(J2000)} &
  \multicolumn{1}{c|}{BaF2$\pm$e\_BaF2} &
  \multicolumn{1}{c|}{Silica$\pm$e\_Silica} &
  \multicolumn{1}{c|}{B15$\pm$e\_B15} &
  \multicolumn{1}{c|}{B13$\pm$e\_B13} &
  \multicolumn{1}{c|}{B4$\pm$e\_B4} &
  \multicolumn{1}{c|}{G$\pm$e\_G} &
  \multicolumn{1}{c|}{PMRA} &
  \multicolumn{1}{c|}{PMDEC} &
  \multicolumn{1}{c|}{T$_{\mathrm{eff}}$}\\
  
 \multicolumn{1}{|c|}{ } &
  \multicolumn{1}{c|}{hh:mm:ss.ss} &
  \multicolumn{1}{c|}{dd:mm:ss.ss} &
  \multicolumn{1}{c|}{ABmag} &
  \multicolumn{1}{c|}{ABmag} &
  \multicolumn{1}{c|}{ABmag} &
  \multicolumn{1}{c|}{ABmag} &
  \multicolumn{1}{c|}{ABmag} &
  \multicolumn{1}{c|}{ABmag} &
  \multicolumn{1}{c|}{mas/yr} &
  \multicolumn{1}{c|}{mas/yr} & 
  \multicolumn{1}{c|}{K} \\
\hline
HB01 & 23:08:22.18 & -15:36:48.65 & 20.05$\pm$0.18 & 20.27$\pm$0.24 & 19.83$\pm$0.38 & 19.60$\pm$0.03 &	19.40$\pm$0.03 & 17.847$\pm$0.003 & 1.090 & -1.951 & 8,750 \\
HB02 & 23:08:23.34 & -15:36:49.47 & 20.38$\pm$0.22 & 19.86$\pm$0.21 & 20.38$\pm$0.49 & 19.65$\pm$0.02 &	19.47$\pm$0.02 & 17.863$\pm$0.004 & 0.559 & -2.512  & 8,750 \\
HB03 & 23:08:26.14 & -15:36:10.97 &	19.95$\pm$0.23 & 20.18$\pm$0.23 & 19.45$\pm$0.32 & 19.56$\pm$0.03 &	19.54$\pm$0.03 & 17.813$\pm$0.003 &	0.568 & -2.559 & 8,750 \\
HB04 & 23:08:28.05 & -15:37:01.95 &	21.26$\pm$0.35 & 19.38$\pm$0.17	& 20.14$\pm$0.44 & 19.56$\pm$0.02 & 19.43$\pm$0.04 & 17.538$\pm$0.002 & 1.019 &	-2.295 & 8,000 \\
HB05 & 23:08:18.99 & -15:35:51.23 &	20.15$\pm$0.20 & 19.22$\pm$0.13	& 19.30$\pm$0.30 & 19.42$\pm$0.03 &	19.43$\pm$0.03 & 18.045$\pm$0.002 & 0.481 &	-2.140 & 9,250 \\
HB06 & 23:08:37.71 & -15:38:56.00 & 19.47$\pm$0.19 & 19.66$\pm$0.20	& 19.88$\pm$0.39 & 19.44$\pm$0.02 & 19.32$\pm$0.03 & 17.985$\pm$0.002 & 0.504 & -2.417 & 9,000 \\
HB07 & 23:08:20.46 & -15:37:35.75 & 20.21$\pm$0.22 & 19.18$\pm$0.20 & 19.56$\pm$0.34 & 19.43$\pm$0.02 & 19.36$\pm$0.03 & 17.890$\pm$0.002 & 0.801 & -2.432 & 9,000 \\
HB08 & 23:08:36.71 & -15:36:35.41 & 19.62$\pm$0.14 & 19.75$\pm$0.22 & 18.81$\pm$0.24 & 19.26$\pm$0.03 & 19.32$\pm$0.02 & 17.735$\pm$0.003 & 0.372 & -2.306 & 9,000 \\
HB09 & 23:08:31.51 & -15:36:33.18 & 20.13$\pm$0.22 & 19.17$\pm$0.12 & 19.37$\pm$0.31 & 19.31$\pm$0.02 & 19.39$\pm$0.03 & 18.019$\pm$0.002 & 1.135 & -2.663 & 9,250 \\
HB10 & 23:08:19.08 & -15:36:34.00 & 19.40$\pm$0.15 & 19.01$\pm$0.16 & 18.97$\pm$0.26 & 19.26$\pm$0.02 & 19.23$\pm$0.02 & 18.338$\pm$0.004 & 0.901 & -2.199 & 10,250 \\
HB11 & 23:08:31.27 & -15:37:32.95 & 19.45$\pm$0.22 & 19.40$\pm$0.24	& 19.57$\pm$0.34 & 19.36$\pm$0.02 & 19.30$\pm$0.02 & 18.127$\pm$0.003 & 1.217 & -1.480 & 9,750 \\
HB12 & 23:08:30.62 & -15:37:10.02 & 19.51$\pm$0.16 & 19.17$\pm$0.16 & 19.20$\pm$0.29 & 19.33$\pm$0.02 & 19.30$\pm$0.03 & 18.237$\pm$0.003 & 0.487 & -2.855 & 9,750\\
HB13 & 23:08:25.66 & -15:37:06.65 & 19.69$\pm$0.18 & 19.69$\pm$0.22 & 19.36$\pm$0.31 & 19.28$\pm$0.02 & 19.27$\pm$0.03 & 17.664$\pm$0.002 & 1.309 & -2.250 & 9,000 \\
HB14 & 23:08:20.75 & -15:37:19.81 & 20.52$\pm$0.27 & 19.71$\pm$0.20 & 19.28$\pm$0.30 & 19.52$\pm$0.02 & 19.52$\pm$0.03 & 17.580$\pm$0.002 & 0.942 & -2.042 & 8,250 \\
HB15 & 23:08:22.47 & -15:36:22.64 & 19.88$\pm$0.19 & 19.44$\pm$0.25 & 18.91$\pm$0.25 & 19.41$\pm$0.02 & 19.53$\pm$0.02 & 18.179$\pm$0.003 & 0.733 & -2.144 & 9,500 \\
HB16 & 23:08:38.19 & -15:33:30.10 & 20.00$\pm$0.19 & 19.36$\pm$0.18 & 19.40$\pm$0.31 & 19.41$\pm$0.02 & 19.39$\pm$0.03 & 17.870$\pm$0.003 & 1.082 & -1.945 & 9,000 \\
HB17 & 23:08:26.23 & -15:38:40.55 & 20.05$\pm$0.21 & 19.39$\pm$0.20 & 19.38$\pm$0.31 & 19.51$\pm$0.03 & 19.38$\pm$0.02 & 17.938$\pm$0.003 & 0.647 & -2.198 & 9,000 \\
HB18 & 23:08:22.56 & -15:38:25.47 & 20.54$\pm$0.21 & 19.96$\pm$0.21 & 19.18$\pm$0.28 & 19.54$\pm$0.03 & 19.54$\pm$0.03 & 17.827$\pm$0.002 & 0.654 & -1.812 & 8,500 \\
HB19 & 23:08:28.96 & -15:36:14.96 & 19.98$\pm$0.26 & 20.52$\pm$0.33 & 19.89$\pm$0.40 & 19.44$\pm$0.02 & 19.43$\pm$0.02 & 17.92$\pm$0.002 & 0.283 & -2.807 & 8,750 \\
HB20 & 23:08:31.30 & -15:37:01.57 & 19.70$\pm$0.15 & 19.37$\pm$0.14 & 19.13$\pm$0.28 & 19.21$\pm$0.02 & 19.20$\pm$0.02 &	18.078$\pm$0.003 & 0.310 &  -2.529 & 9,500 \\
HB21 & 23:08:29.22 & -15:35:14.88 & 19.73$\pm$0.147 & 19.33$\pm$0.18 & 19.06$\pm$0.27 & 19.35$\pm$0.02 & 19.33$\pm$0.03 & 	18.026$\pm$0.002 & 0.521 & -2.473 & 9,500 \\
HB22 & 23:08:26.65 & -15:36:07.52 & 19.59$\pm$0.16 & 19.23$\pm$0.17 & 19.10$\pm$0.28 & 19.20$\pm$0.02 & 19.29$\pm$0.03 & 18.373$\pm$0.004 & 1.222 & -2.103 & 10,500 \\
HB23 & 23:08:27.28 & -15:37:05.99 & 21.32$\pm$0.41 & 20.12$\pm$0.31 & 19.40$\pm$0.31 & 19.65$\pm$0.02 & 19.68$\pm$0.04 & 17.638$\pm$0.003 & 0.631 & -2.472 & 8,000 \\
HB24 & 23:08:26.82 & -15:36:34.66 & 20.04$\pm$0.22 & 19.59$\pm$0.23 & 19.20$\pm$0.29 & 19.27$\pm$0.03 & 19.29$\pm$0.03 & 17.940$\pm$0.003 & 1.173 & -2.125 & 9,000 \\
HB25 & 23:08:27.27 & -15:37:20.24 & 19.77$\pm$0.19 & 19.77$\pm$0.26 & 19.32$\pm$0.30 & 19.30$\pm$0.02 & 19.32$\pm$0.02 & 18.114$\pm$0.003 & 0.569 & -2.321 & 9,500 \\
HB26 & 23:08:28.40 & -15:35:46.21 & 19.82$\pm$0.16 & 20.12$\pm$0.32 & 19.30$\pm$0.30 & 19.51$\pm$0.02 & 19.36$\pm$0.02 & 17.808$\pm$0.007 & 0.689 & -2.669 & 8,750 \\
HB27 & 23:08:30.99 & -15:37:45.53 & 19.58$\pm$0.119 & 18.90$\pm$0.14 & 18.90$\pm$0.25 & 19.16$\pm$0.02 & 19.25$\pm$0.03 & 18.435$\pm$0.003 & 1.676 & -2.512 & 10,500 \\
HB28 & 23:08:26.65 & -15:38:06.14 & 20.08$\pm$0.23 & 19.85$\pm$0.23 & 19.41$\pm$0.32 & 19.33$\pm$0.02 & 19.39$\pm$0.03 & 17.672$\pm$0.002 & 0.860 & -2.543 & 8,750 \\
HB29 & 3:08:29.73 & -15:35:00.16 & 19.94$\pm$0.20 & 19.64$\pm$0.24 & 19.19$\pm$0.27 & 19.32$\pm$0.02 & 19.35$\pm$0.02 & 18.034$\pm$0.003 & 0.392 & -2.437 & 9,250 \\
HB30 & 23:08:25.38 & -15:36:39.99 & 20.43$\pm$0.27 & 19.64$\pm$0.18 & 19.39$\pm$0.31 & 19.39$\pm$0.03 & 19.37$\pm$0.03 & 17.932$\pm$0.003 & 0.159 & -2.411 & 8,750 \\
HB31 & 23:08:22.22 & -15:36:00.59 & 19.58$\pm$0.21 & 18.93$\pm$0.14 & 19.31$\pm$0.30 & 19.33$\pm$0.02 & 19.32$\pm$0.02 & 18.244$\pm$0.003 & 0.678 & -2.222 & 10,000 \\
HB32 & 23:08:22.88 & -15:36:50.52 & 21.08$\pm$0.38 & 20.42$\pm$0.27 & 18.94$\pm$0.25 & 19.74$\pm$0.03 & 19.45$\pm$0.02 & 17.547$\pm$0.002 & 1.514 & -2.542 & 8,000 \\
HB33 & 23:08:27.96 & -15:36:17.76 & 19.76$\pm$0.19 & 19.70$\pm$0.21 & 19.72$\pm$0.37 & 19.36$\pm$0.02 & 19.38$\pm$0.03 & 17.938$\pm$0.002 & 1.082 & -2.199 & 9,000 \\
HB34 & 23:08:34.97 & -15:38:21.46 & 20.39$\pm$0.24 & 19.91$\pm$0.27 & 19.51$\pm$0.33 & 19.48$\pm$0.02 & 19.44$\pm$0.03 & 17.922$\pm$0.002 & 0.593 & -2.091 & 8,750 \\
HB35 & 23:08:22.13 & -15:36:10.09 & 20.13$\pm$0.19 & 19.43$\pm$0.21 & 19.44$\pm$0.32 & 19.43$\pm$0.03 & 19.38$\pm$0.02 & 18.150$\pm$0.003 & 0.632 & -2.659 & 9,250 \\
HB36 & 23:08:27.00 & -15:35:48.20 & 20.22$\pm$0.23 & 19.12$\pm$0.14 & 19.28$\pm$0.30 & 19.23$\pm$0.03 & 19.36$\pm$0.03 & 18.255$\pm$0.005 & - & - & 9,500 \\
HB37 & 23:08:26.30 & -15:37:17.49 & 19.93$\pm$0.20 & 18.90$\pm$0.17 & 19.08$\pm$0.27 & 19.20$\pm$0.03 & 19.30$\pm$0.04 & 18.370$\pm$0.006 & 0.358 & -3.855 & 10,000 \\
HB38 & 23:08:30.95 & -15:36:56.17 & 20.54$\pm$0.22 & 19.94$\pm$0.24 & 19.58$\pm$0.34 & 19.41$\pm$0.03 & 19.44$\pm$0.02 & 17.732$\pm$0.003 & 0.734 & -2.042 & 8,500 \\
HB39 & 23:08:25.85 & -15:35:30.37 & 19.83$\pm$0.17 & 19.71$\pm$0.21 & 19.06$\pm$0.27 & 19.31$\pm$0.02 & 19.33$\pm$0.02 & 17.852$\pm$0.003 & - & - & 9,000 \\
EHB01 & 23:08:25.10 & -15:36:26.86 & 20.28 $\pm$ 0.20 & 19.84 $\pm$ 0.23 & 20.17 $\pm$ 0.45 & 20.19 $\pm$ 0.04 & 20.54 $\pm$ 0.05 & - & - & - & 29,000 \\

\hline\end{tabular}

 }
\end{table*}

\subsection{Blue Horizontal Branch Stars (BHBs)}
The UV, UV-optical and optical CMDs for all the 39 BHBs are shown in \autoref{fig:uvcmd}. Out of 39 BHBs, two are not having proper motion information but they are included in the list based upon their position in CMDs (light blue solid circles in \autoref{fig:uvcmd}). BHBs are the hottest and brightest sources in UV, therefore, only these sources along with the lone EHB star are detected in FUV. That's why no other sources are seen in Silica$-$NUVB4 vs Silica CMD. Whereas in other CMDs, we see all detected sources including BHBs. For all BHBs in Silica$-$NUVB4 vs Silica CMD, the magnitude in Silica filter ranges from 19.10 to 20.51 mag and the Silica$-$NUVB4 color ranges from $-$0.33 to 1.03 mag (\textit{upper left panel}, \autoref{fig:uvcmd}). Of the 39 BHBs of the cluster, 35 are common in the {\em GALEX} observation \citep{Schiavon2012}. We found 11 BHBs of NGC 7492 in SIMBAD, out of which 10 were reported as blue stars from the {\em GALEX} observation \citep{Atlee2007} and one was reported as field BHB star by \cite{Christlieb2005}, which we verified to be a cluster member based upon its proper motion and position in CMDs. The early studies by \cite{Buonanno1987, Cote1991} reported a less number of BHBs as their observations were concentrated within the half light radius of the cluster. We have observed 18 BHBs out of 39 lie outside the half light radius of the cluster. The photometric details of the 39 BHBs observed with the five UVIT filters are given in \autoref{tab:BHB_catalog_apendix}. 

\subsection{Red Giant Branch Stars (RGBs)}
The morphology of RGBs in UV-optical CMDs \citep{Schiavon2012, Piotto2015, Subramaniam2016, sahu288} are different from IR/optical CMDs \citep{Sarajedini2007, Riffel2011, Vanderbeke2014, Cohen2015}. The sources which are brighter but not bluer than BHBs in IR/optical get fainter and redder in UV and UV-optical CMDs as we move towards optical to UV filters. Since the cluster is located at a large distance, the RGBs become very faint to observe in FUV bands. Therefore, we do not see the RGBs in the Silica$-$NUVB4 vs Silica CMD. In the NUVB13$-$NUVB4 vs NUVB13 CMD, we see that RGBs are distributed throughout the color range from $-$0.1 to 1.2 mag towards the fainter magnitudes of 22 to 23 mag. Therefore, we identified the NUV detected RGBs based upon the positions of their optical counterparts in the NUVB4$-$G vs NUVB4 and BP$-$RP vs BP CMDs (\textit{red dots in lower right panel} in \autoref{fig:uvcmd}), where RGBs are well separated from AGBs and SGBs. We chose to keep clean RGBs up to 19.2 mag in the BP band. We found 91 RGBs in the NUVB4 and 53 RGBs in NUVB13 down to 23.0 mag. The range of color for RGBs in NUVB4$-$G vs NUVB4 CMD is from 2.29 to 6.50 mag, whereas the NUVB4 magnitude ranges from 21.11 to 22.63 mag. Similarly, the value of color for the RGBs ranges from 0.32 to 0.92 mag whereas the magnitude ranges from 15.70 to 19.26 mag in BP$-$RP vs BP CMD. We have detected 6 AGBs in both the NUVB4 and NUVB13 filters, 28 SGBs in NUVB4 filter, 6 AGBs and 6 SGBs in NUVB13 filter. These sources are plotted on NUV$-$NUV, NUV-optical and optical CMDs in \autoref{fig:uvcmd}. The AGBs are shown with upper solid triangles and SGBs are shown with gray lower solid triangles. We see that AGBs are clearly separated from other branch sources and are lying above the RGBs in both the NUV$-$NUV and NUV-optical CMDs.  

\subsection{Variable Stars}
We cross-matched the UVIT observed sources with the catalog of variable stars provided by \cite{Clement2001, Clement2017}\footnote{\href{http://www.astro.utoronto.ca/~cclement/cat/C2305m159} {http://www.astro.utoronto.ca/~cclement/cat/C2305m159}} for NGC 7492. We detected five variable stars (three RR Lyrae stars and two SXPhes) in NUV filters and one in both FUV and NUV filters. All these variable stars are shown with dark orange diamonds in the CMDs (\autoref{fig:uvcmd}). Out of the five variable stars, two RR Lyrae stars and two SXPhes lie in the inner part of the cluster and the fifth RR Lyrae star which is observed in both NUV and FUV bands lies outside the half light radius of the cluster. Of the three RR Lyrae stars, two are situated in the instability strip (the gap between BHBs an RGBs) and the other one lies within the BHB region in all the CMDs (\autoref{fig:uvcmd}). Since the UVIT telescope observes in photon counting mode, it is possible to study the variability and generate light curves of short period variables \citep{Subramaniam2017}.

\subsection{Blue Straggler Stars (BSs)}
In NUVB4$-$G vs NUVB4 CMD (\autoref{fig:uvcmd}, \textit{lower left panel}), we do not find a clear separation of the MS turn-off region from the SGBs. Hence, to avoid confusion while selecting BSs, we chose only those sources which were clearly above the MS turn-off bump. We have identified only four BSs which were bright enough in NUV filters to be separated out from MS-turnoff and SGBs in the NUV$-$optical CMD. However, we do not find any BSs in FUV filters due to its lower detection limit (21.5 mag) towards the fainter sources. In this cluster, \cite{Cote1991} have reported 27 BSs with optical photometry. The radial distribution of BSs in Figure 13 of \cite{Cote1991} and \autoref{fig:mag_dist} of this paper (see \autoref{sec:discussion}) show that the UVIT detected BSs are lying up to 1.5$'$ from the center of the cluster.

\section{Constraining various parameters of the cluster}
\label{sec:param}

We have cross-matched the UVIT detected sources of NGC 7492 with the catalog of ground-based observations provided by \citet{Stetson2019}. In \autoref{fig:cmd_ground}, we have plotted the V$-$I vs V CMD for all the sources obtained from the wide-field ground-based observations in gray solid circles and over-plotted the UVIT detections in magenta solid circles. The presence of stars of almost all possible stages of evolution are noticed in the ground-based observation whereas only the SGBs and the stars of later stages of evolution are seen in the UVIT observations. Various parameters of the cluster such as metallicity, age, distance, etc., are generally estimated by fitting isochrones on the turn-off region of the main-sequence (MSTOS) and on the SGBs. In \autoref{fig:cmd_ground}, we see a sharp turn-off region along with good number of SGBs which we have used in constraining the cluster parameters.

In the following subsections, we have constrained distance to the cluster and He-abundance of BHBs using BaSTI model isochrones and zero-age horizontal branch (ZAHB) over UV, UV$-$optical and optical CMDs.

\subsection{Distance estimation}

\begin{figure*}
 \centering
    \subfloat[\label{age}]{\includegraphics[]{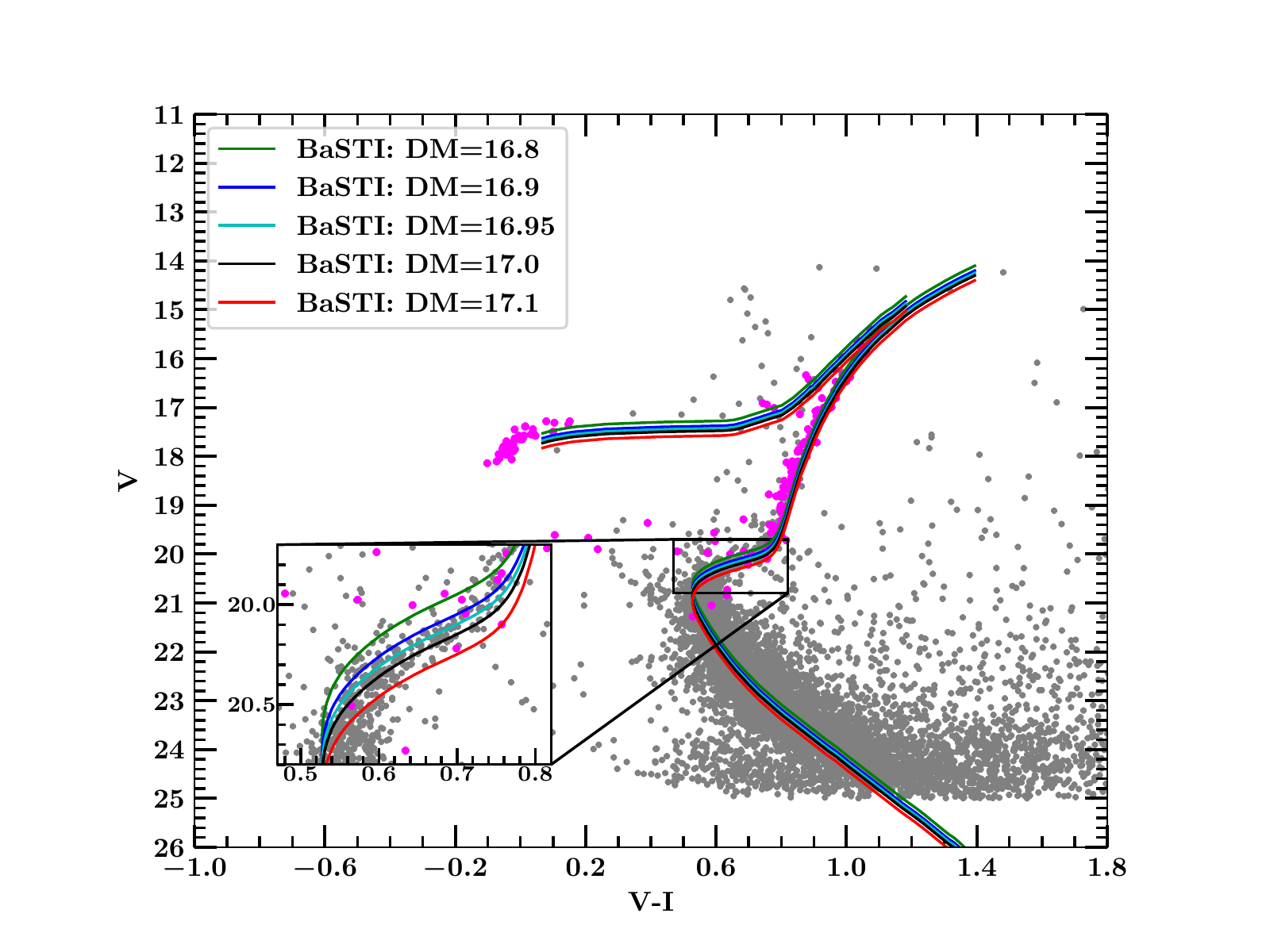}} 
    \caption{V-I vs V CMD of ground-based photometry of NGC 7492 provided by \citet{Stetson2019} (gray solids)). The sources cross-matched with UVIT detections are over-plotted in magenta solids. The BaSTI isochrones generated for age 12.0 Gyr and $[Fe/H] = -1.8$ have been over-plotted for various distance modulus (DM) as indicated in the legend. The inset panel shows the SGB region.}
    \label{fig:cmd_ground}
\end{figure*}

We have generated isochrones of the updated BaSTI stellar evolution models \citep{Hidalgo2018} with the following input parameters: $[Fe/H]=-1.8$, age = 12.0 Gyr and distance-modulus (DM) values ranging from 16.8 to 17.1. We show the over-plotted isochrones on V-I vs V CMD in \autoref{fig:cmd_ground}. The inset plot in  \autoref{fig:cmd_ground} shows the SGB region of the cluster. The best fitted isochrone to the SGB region is having a distance modulus 16.95 (cyan line). The isochrones get merged below the turn-off point and do not show much variation in the MS or RGB region. The isochrones at HB region also show a clear deviation at the distance modulus of 17.1. Thus, we conclude that the distance modulus of the cluster to be 16.95$\pm$0.05 which is equivalent to 24.5$\pm$0.5 kpc. Previously, \citet{Figuera2013} have estimated the distance to the cluster using period-luminosity relation of variable stars. They found it to be 25.2$\pm$1.8 kpc using SxPhes stars and 24.3 $\pm$ 0.5 kpc using RR Lyrae stars which are in agreement within error with our estimation.

\subsection{Helium abundance of BHBs}
\label{He_enrich}

\begin{figure*}
  \centering
    \subfloat[\label{He_VI}]{\includegraphics[width=0.33\textwidth]{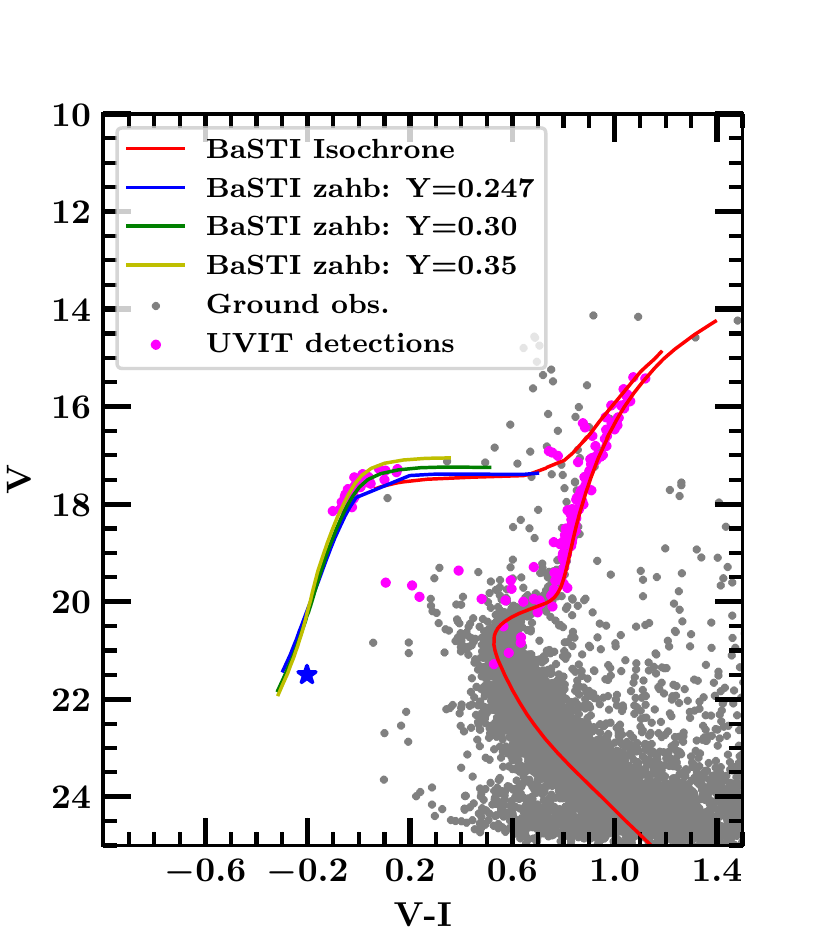}} 
    \subfloat[\label{He_B4}]{\includegraphics[width=0.33\textwidth]{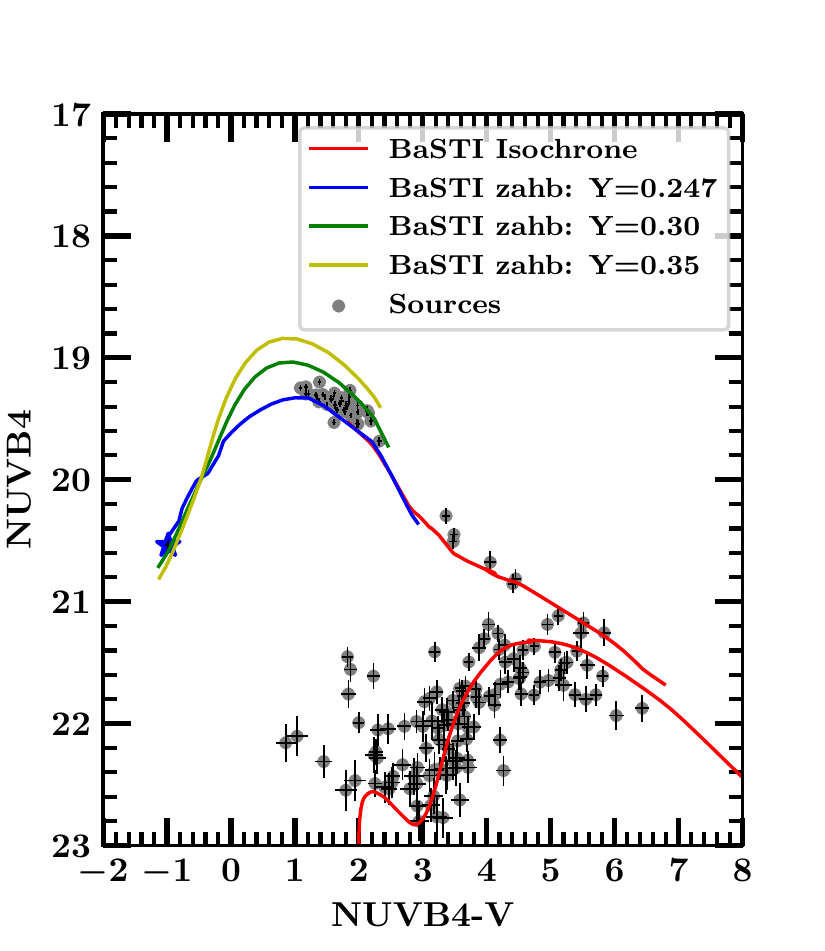}}
    \subfloat[\label{He_B13}]{\includegraphics[width=0.33\textwidth]{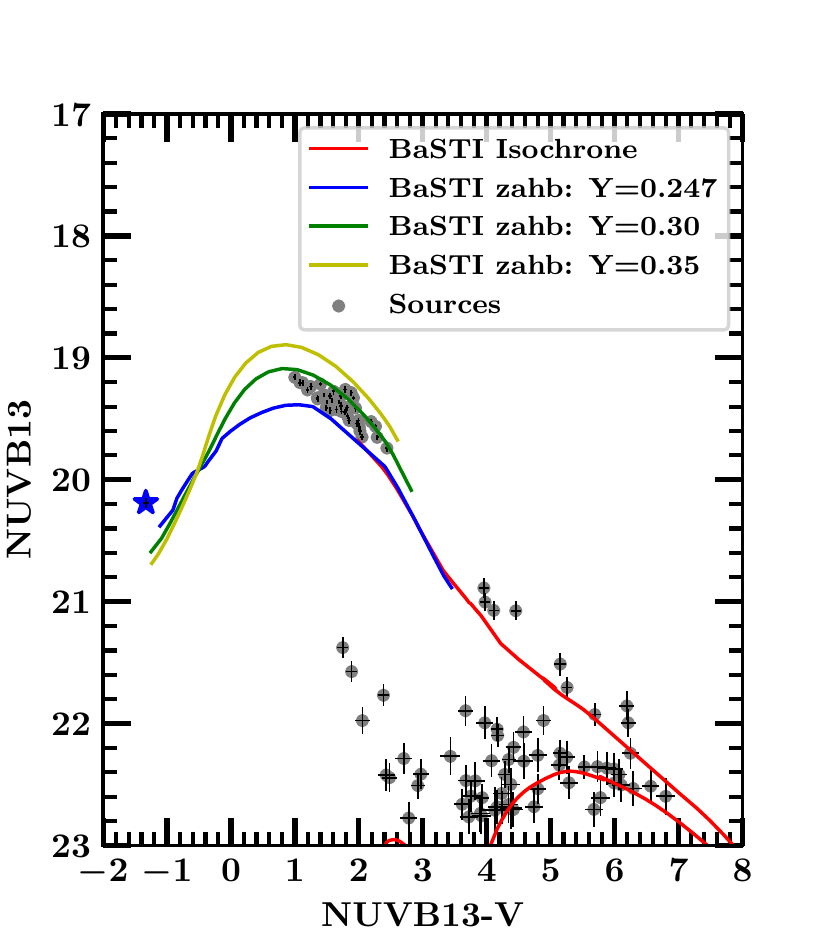}} \\
  \centering    
    \subfloat[\label{He_B15}]{\includegraphics[width=0.33\textwidth]{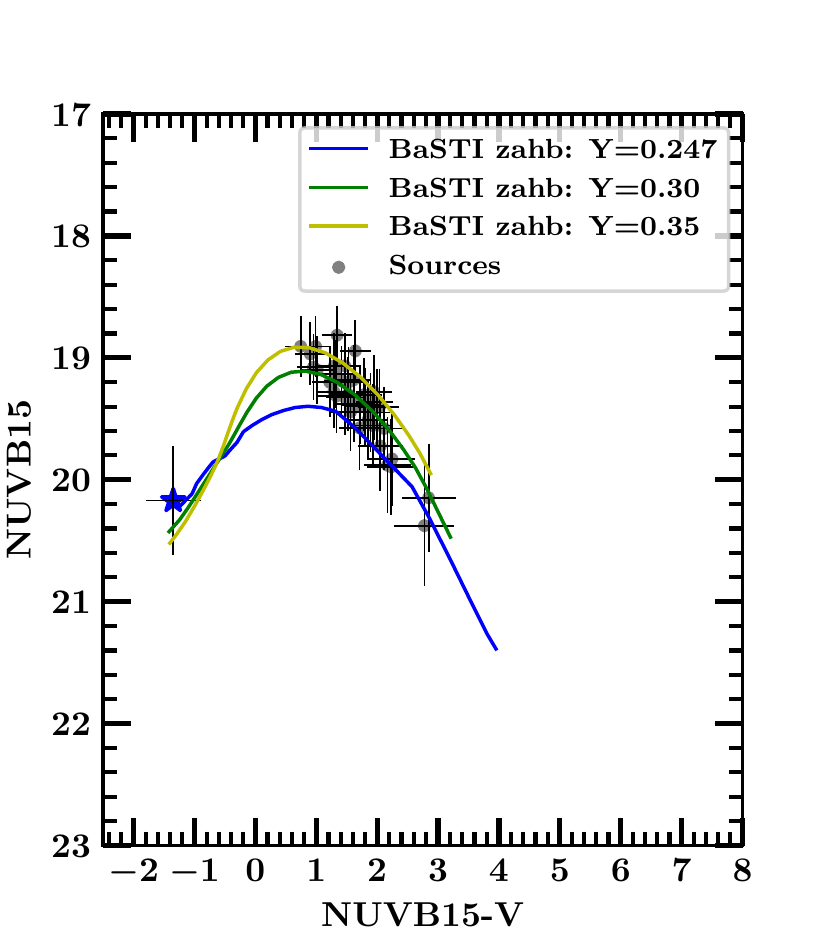}}
    \subfloat[\label{He_sil}]{\includegraphics[width=0.33\textwidth]{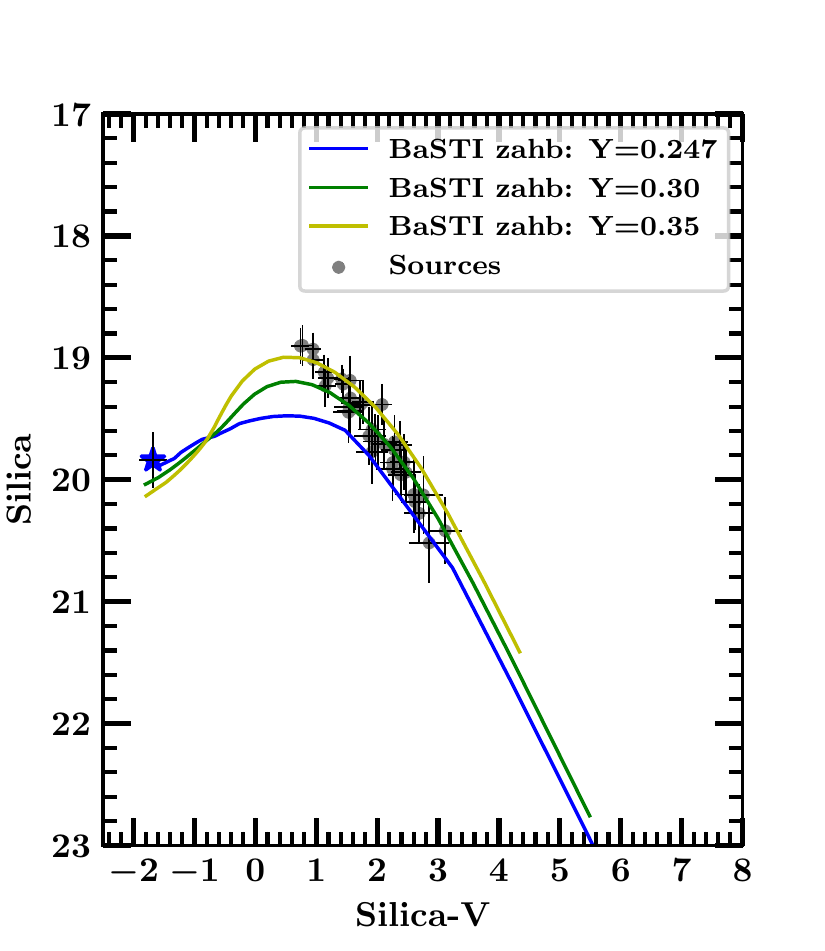}}
    \subfloat[\label{He_baf2}]{\includegraphics[width=0.33\textwidth]{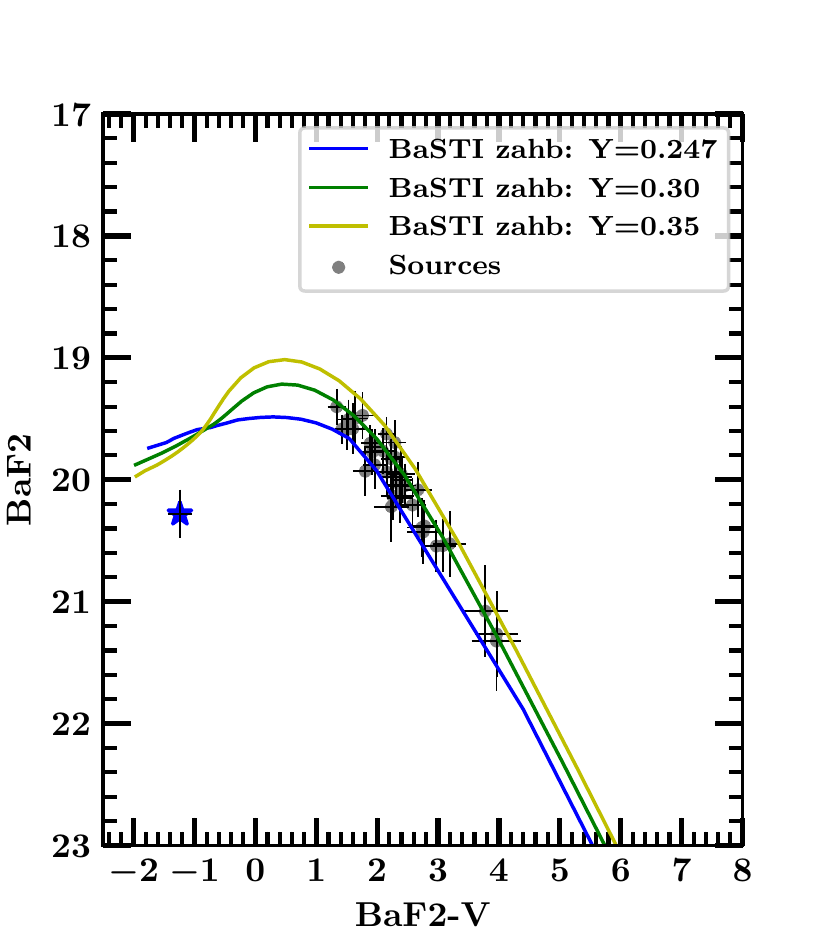}} 
  
    \caption{Optical, NUV$-$optical and FUV$-$optical CMDs of UVIT sources are shown along with BaSTI isochrones with He-abundant ZAHB for He-abundance 0.247 (blue line), 0.300 (green line) and 0.350 (yellow line). In panel (a) we have plotted the V$-$I vs V CMD of all the sources obtained from ground-based observations in gray solids. Cross-matched UVIT sources are overploted in magenta solids. In panels (b), (c), (d), (e), and (f), we have plotted CMDs of UVIT filters NUVB4, NUVB13, NUVB15, Silica, and BaF2, respectively in combination with ground-based observations of V filter. The UVIT detected sources in all the CMDs are shown in black solids along with the associated error-bars. The lone EHB star is marked in blue asterisk in all the CMDs.} 
    \label{fig:cmd_he}
\end{figure*} 

The temperature of UVIT observed BHBs ranges from 8,000 K to 10,500 K (see \autoref{sec:bhb}), which is suitable to see the variation of He-abundances in them \citep{Marino2014}. In \autoref{fig:cmd_he}, we have plotted the $\alpha$-enhanced ZAHB from the updated BaSTI model for three He-abundances, Y $=$ 0.247, 0.300 and 0.350, and $[Fe/H]=-1.8$ over optical, NUV$-$optical and FUV$-$optical CMDs. We see He-abundances of BHBs, Y in the range, 0.247 to 0.350 in V$-$I vs V, Silica$-$V vs Silica, NUVB15$-$V vs NUVB15, and BaF2$-$V vs BaF2 CMDs (Figures \ref{He_VI}, \ref{He_sil}, \ref{He_B15}, and \ref{He_baf2}, respectively). Similarly, it is between Y = 0.247 and 0.300 in NUVB13$-$V vs NUVB13 and NUVB4$-$V vs NUVB4 CMDs (Figures \ref{He_B13} and \ref{He_B4}, respectively). \citet{Marino2014, Villanova2012} and \citet{Behr2003} have used the prominent He I line at 5875 \AA\ to estimate the He-content of BHBs in several GGCs. Since this line is lying within the waveband of V filter ($\lambda_{\mathrm{eff}}$ = 5445 \AA), we also see He enhancement in BHBs of the cluster NGC 7492 (Figure \autoref{He_VI}). The NUVB15 filter ($\lambda_{\mathrm{eff}}$ = 2193 \AA) is heavily affected by the extinction bump ($\lambda_{\mathrm{eff}}$ = 2175 \AA) and also the signal to noise ratio was very low due to small exposure time for BHBs observed in this filter. We do not consider this filter as a good estimator of He-content though we see a He-enrichment of 0.05 to 0.10  (Figure \autoref{He_B15}). From the Figures \ref{He_sil} and \ref{He_baf2}, it is obvious that the FUV filters are the best estimator of the He-abundances of BHBs and we see a clear variation from Y = 0.247 to Y = 0.350 in the FUV filters. However, it is very difficult to comment on the contribution of prominent He lines within the wavelength range of our FUV filters since the synthetic spectra of BHBs with He-abundance higher than Y = 0.247 is not available in FUV regime. Since, we have used $\alpha$-enhanced plus He-normal and also He-enhanced ZAHB, we suggest that the enhancement in metal content (or He) are definitely present in BHBs of NGC 7492.

Earlier the He-content of the cluster was estimated to be 0.23$\pm$0.06 using R-parameter, which is the number ratio of BHBs to upper RGBs in B$-$V vs V CMD \citep{Cote1991,Buonanno1987}. While calculating the He-content, they used only 18 BHBs, 20 RGBs and 5 AGBs as their observations were concentrated within the core of the cluster. With the observation up to the tidal radius of the cluster, we have detected more number of these sources (i.e., 39 BHBs, 20 RGBs and 6 AGBs) in the cluster than previously observed. Using the number of these sources in the formulas given in \citet{Buzzoni1983}, we have calculated R, and R$'$ and hence, the corresponding He abundances as follows:
\begin{equation}
R = \frac{N(HB)}{N(RGB)} = \frac{39}{20} = 1.95
\end{equation}
% and 
\begin{equation}
    R' = \frac{N(HB)}{N(RGB+AGB)} = \frac{39}{20+6} = 1.5
\end{equation}
%\\
\begin{equation}
    Y=0.380\times \log(R)+0.176 = 0.380\times \log(1.95)+0.176 = 0.286
\end{equation}
%and 
\begin{equation}
    Y'= 0.457\times \log(R')+ 0.204 = 0.457\times \log(1.5)+0.204= 0.284
\end{equation}

\noindent Using the Poisson error in the number statistics of BHBs and RGBs, the estimated He-abundance of the cluster is 0.28$\pm$0.05. Hence, the He-abundance obtained from the CMDs and the R-parameters are in good agreement.

\section{Temperature distribution of BHBs}
\label{sec:bhb}

BHBs are very hot stars with T$_{\mathrm{eff}}$ ranging from 8,000 K to 30,000 K. The T$_{\mathrm{eff}}$ of BHBs mainly depends on the amount of mass loss during the helium flash at the RGB tip. The larger mass loss gives the lesser H-rich envelope and hence, the higher T$_{\mathrm{eff}}$ of BHBs and vice versa. We have estimated T$_{\mathrm{eff}}$ of BHBs using the Kurucz stellar atmospheric model \citep[][hereafter Kurucz model]{kurucz1997,kurucz2004}. However, it is done by two methods; (i) using the color-temperature relation obtained from the Kurucz model and then matching the observed colors with the model generated colors for various UVIT filters (ii) using SED analysis of the observed fluxes obtained from the five UVIT filters, and the photometric fluxes obtained from the archival catalog of various filters of {\em GALEX}, GAIA, Pan-STARRS, CFHT, and CTIO-4m telescopes. We have described the two methods in the following subsections separately.

\subsection{Temperature distribution using color-temperature relation}

\begin{figure}
    \centering
    \includegraphics[width=\columnwidth]{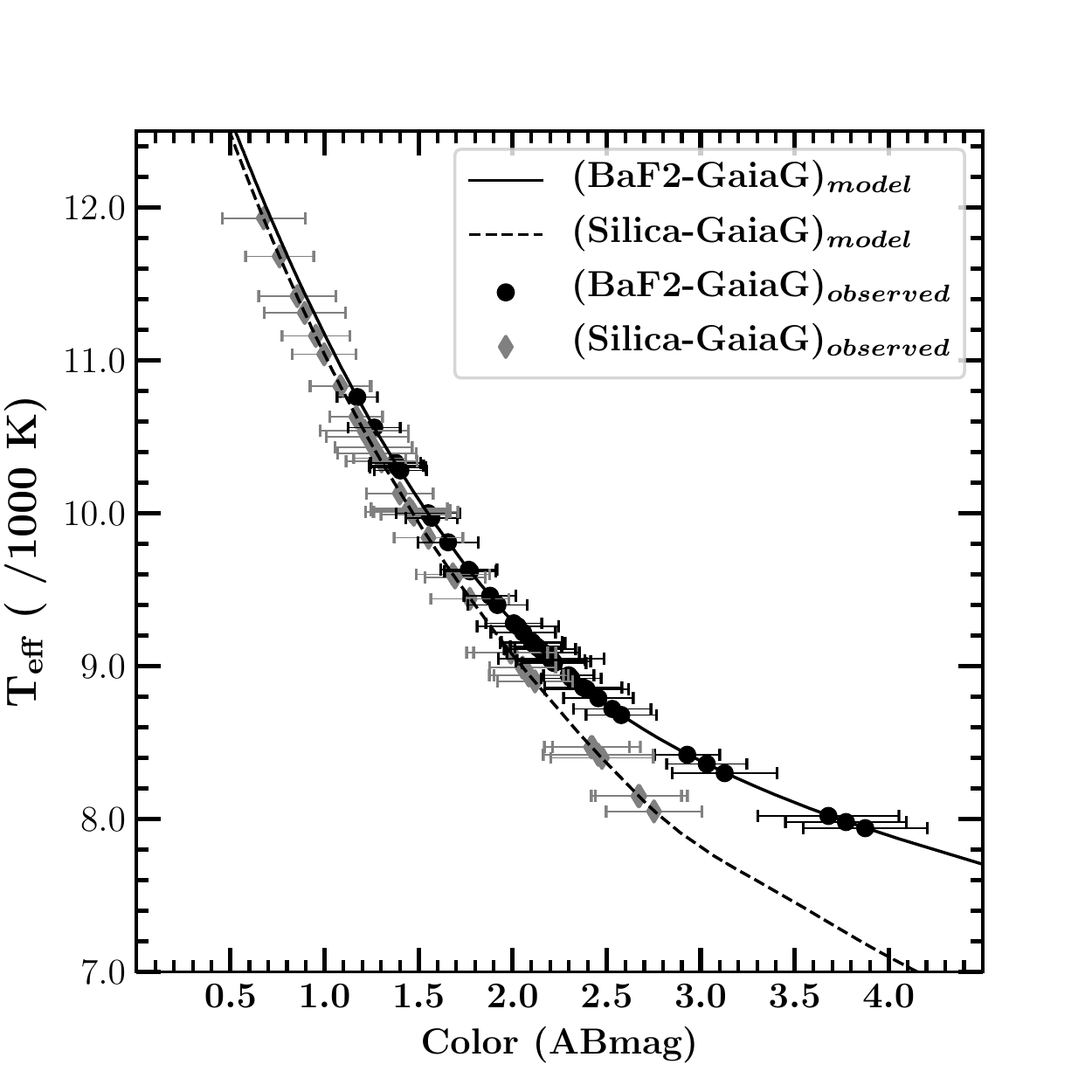}
    \caption{ The color-temperature relations of Silica $-$ GaiaG and BaF2 $-$ GaiaG colors with their corresponding effective temperatures from the Kurucz model. The synthetic color-temperature curves of log(g) = 3.0 in solid (dashed) lines are shown for BaF2 $-$ GaiaG (Silica $-$ GaiaG) colors. The observed colors, BaF2 $-$ GaiaG and Silica $-$ GaiaG are over plotted with black and gray points. The error-bars in the observed colors are the effective error obtained from the magnitude errors of both filters.}
    \label{fig:color-temp}
\end{figure}

We have used Spanish Virtual Observatory \footnote{\href{http://svo2.cab.inta-csic.es/theory/newov2/syph.php}{http://svo2.cab.inta-csic.es/theory/newov2/syph.php} } \citep[SVO,][]{svo2008} to develop synthetic photometry for UVIT filters using the Kurucz model. For a given temperature, metallicity and surface gravity, SVO produces synthetic photometric fluxes for various filters by convolving the filter responses with the Kurucz model.  We generated synthetic fluxes for Silica, BaF2 and Gaia G filters adopting metallicity $[Fe/H]= - 2.0$ dex (nearest grid to the cluster metallicity, $Fe/H] = -1.8$), the temperature in the range from 4,000 K to 50,000 K (with a step-size of 250 K from 4,000 to 13,000 K and a step size of 1,000 K above that) and the value of $log(g) = 3.0$. We interpolated the model generated effective temperatures with a step  size of 10 K using cubic interpolation to get a closer color match for the observed BHBs. In \autoref{fig:color-temp}, we have plotted the model generated color - temperature curves separately for BaF2 $-$ GaiaG and Silica $-$ GaiaG colors and overlaid on the observed colors of all BHBs. We got a range of T$_{\mathrm{eff}}$ from 8,000 K to 12,000 K using Silica $-$ GaiaG colors and 8,000 K to 10,750 K using BaF2 $-$ GaiaG colors for all the BHBs. We found that the observed Silica magnitudes for most of the BHBs are brighter than the corresponding BaF2 magnitudes. This difference in observed magnitudes gives rise to a shifting of Silica $-$ GaiaG color relative to the BaF2 $-$ GaiaG color  ($\Delta_{c} > 0.5$ mag) as seen in \autoref{fig:color-temp}. Again, we see from the figure that there is not much difference between the two synthetic colors at higher T$_{\mathrm{eff}}$ ( $>$ 9,000 K, color $\leq$ 2.0 mag). Therefore, T$_{\mathrm{eff}}$ we obtained from Silica $-$ GaiaG colors is slightly higher than the values obtained from the BaF2 $-$ GaiaG colors.

\subsection{Temperature distribution from SEDs} 
\begin{table*}
    \caption{List of the Telescopes and their filters used in the SEDs fit. }
    \label{tab:telescope}
    \centering
    
    \adjustbox{max width=\textwidth} {
    \begin{tabular}{ c c c c c  }
    \hline
    Telescope & Filters & Wavelength range &  Date of Observation & Reference\\
    \hline
    UVIT & BaF2, Silica, NUVB15, NUVB13, NUVB4 & 1350-2800 \AA & 2016 Oct 19 & This paper \\ 
    {\em GALEX} & FUV, NUV & 1350-3000 \AA & 2005 Aug 26 & \citet{Schiavon2012} \\
    GAIA  & G, BP, RP & 3300-10600 \AA & - & \citet{GaiaCatalog2018} \\
    PAN-STARRS & g, r, i, z, y & 3900-10800 \AA & 2012 June 22 & \citet{Chambers2016} \\
    CTIO-4m  & U, B, V, R, I & 3000-11800 \AA & 2009 Nov 23 & \citet{Stetson2019} \\
    CFHT  & g, r       &  3900-7200 \AA  & 2009 Nov & \citet{Munoz2018} \\

    \hline
    \end{tabular} }
\end{table*}

\noindent The effective temperature of a source depends upon its SED through all the wavelength range. To study SEDs of the detected BHBs, we used UVIT, {\em GALEX}, GAIA, PanSTARSS, ground-based UBVRI photometry from \citet{Stetson2019} catalog, and ground-based g and r band photometry from \citet{Munoz2018}. The filter details, wavelength range, and date of observation are given in \autoref{tab:telescope}. When compared the number of BHBs detected in UVIT with {\em GALEX} we found some discrepancy. Out of 39 UVIT detections of BHBs, two were not reported in \cite{Schiavon2012}. Another four BHBs of UVIT detection are reported as two in {\em GALEX} observation as they were not resolved in {\em GALEX} due to its poor resolution. Ultimately, out of 39 BHBs we have used the best available photometric data of 33 BHBs from {\em {\em GALEX}} observations. Thus, we have used the photometric observations in 22 filters for 33 BHBs and in 20 filters for the rest of the 6 BHBs from FUV to NIR.

We used the virtual observatory of SED analyzer \citep[VOSA,][]{svo2008} to fit the observed photometric fluxes with the Kurucz model, ATLAS9 \citep{kurucz1997,kurucz2004}. VOSA includes various grids (spectra) of Kurucz atmosphere models. We have adopted the Kurucz model grids of metallicity $-2.0$ and $-1.5$ (nearest grids to the cluster metallicity, $Fe/H] = -1.8$) to fit the observed flux of BHBs with the Kurucz model. We have used chi-square minimisation technique to fit the observed flux with the model at the corresponding effective wavelengths of photometric filters. We find that the temperature ranges from 8,000 K to 10,500 K using SEDs fit of the BHBs.

\subsection{Comparison of effective temperatures}
\begin{figure}
    \centering
    \includegraphics[width=\columnwidth]{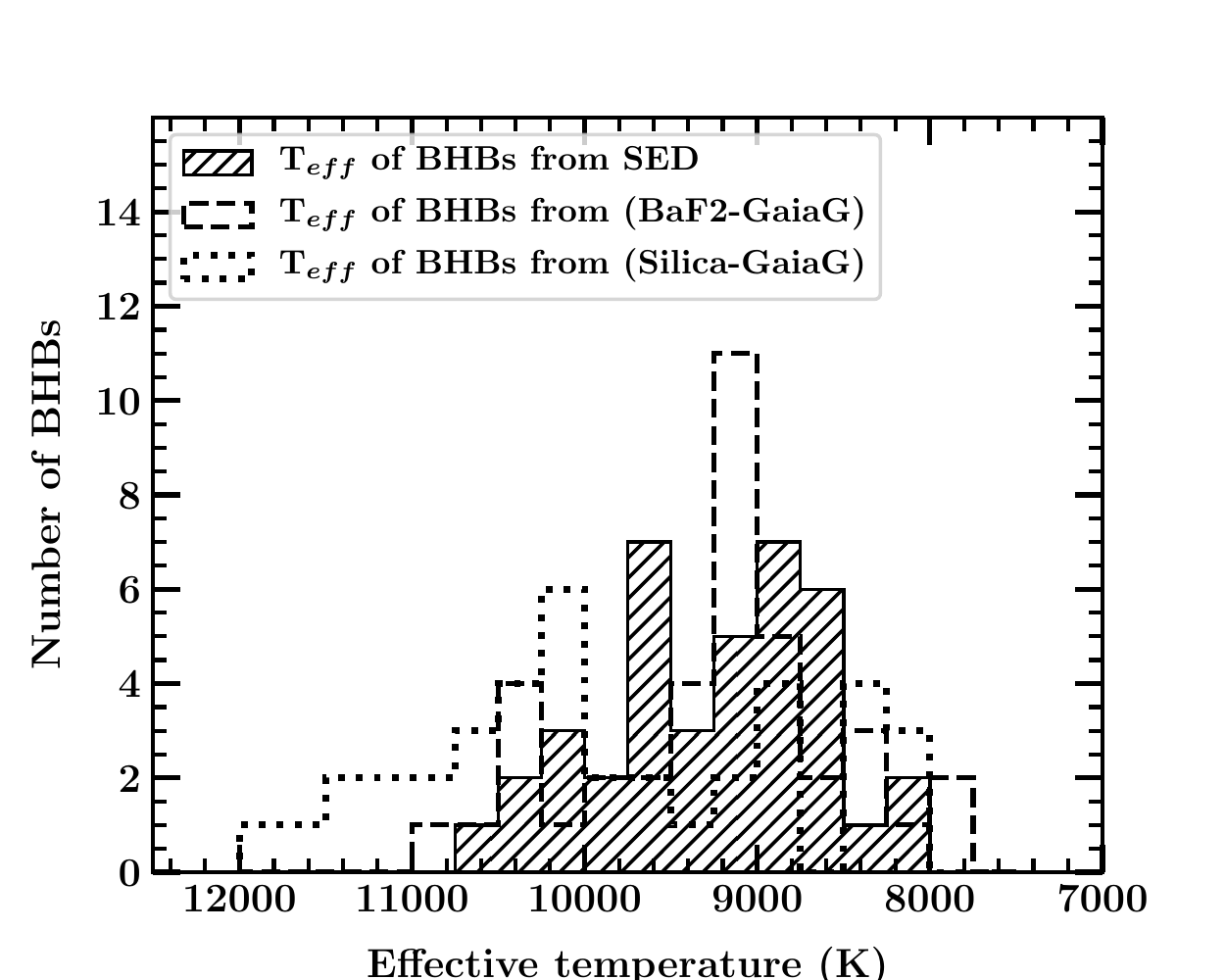}
    \caption{Histogram distributions of effective temperatures of BHBs derived from both SED fitting, and BaF2 $-$ GaiaG and Silica $-$ GaiaG color temperature relations.}
    \label{fig:temp_hist}
\end{figure}

\begin{figure*}
        \centering
        \includegraphics[width=0.498\textwidth]{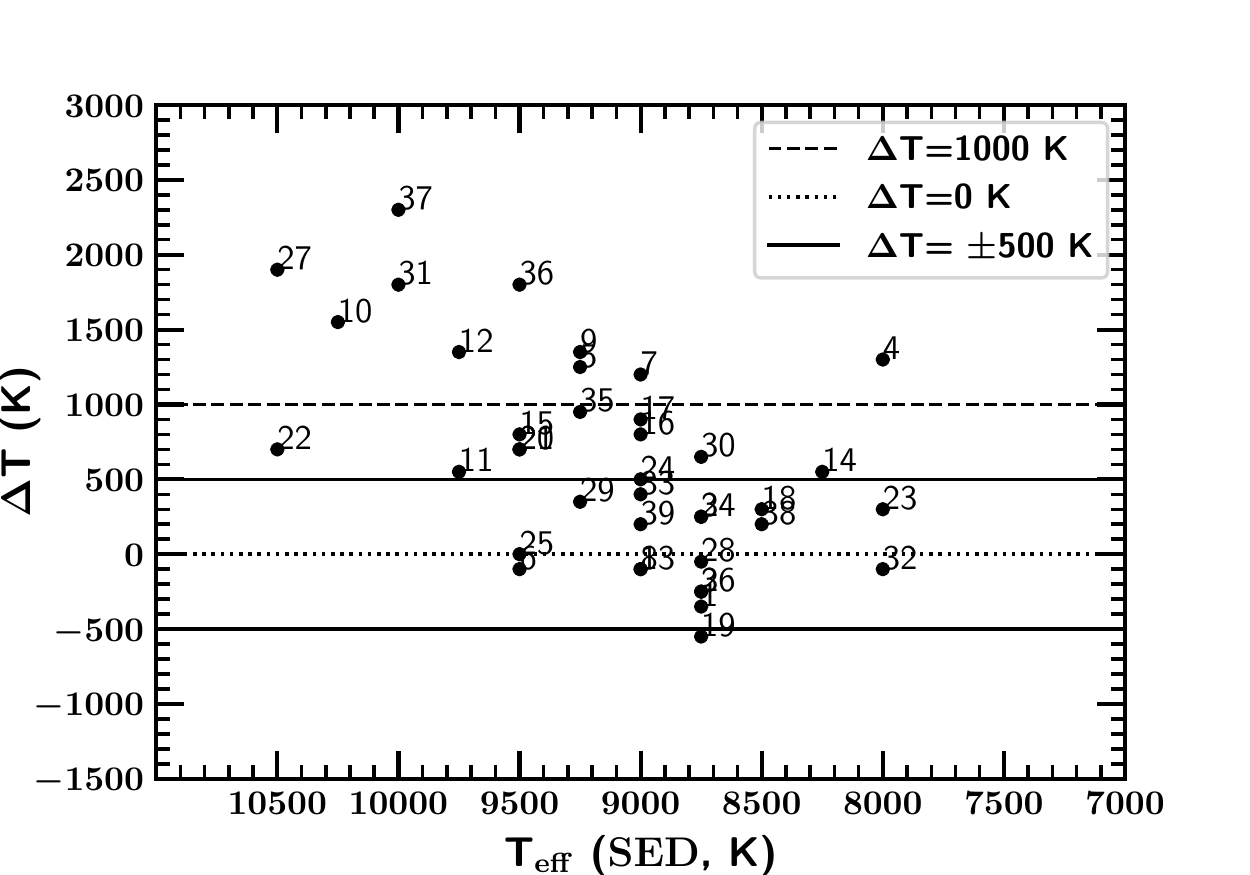}
        \includegraphics[width=0.498\textwidth]{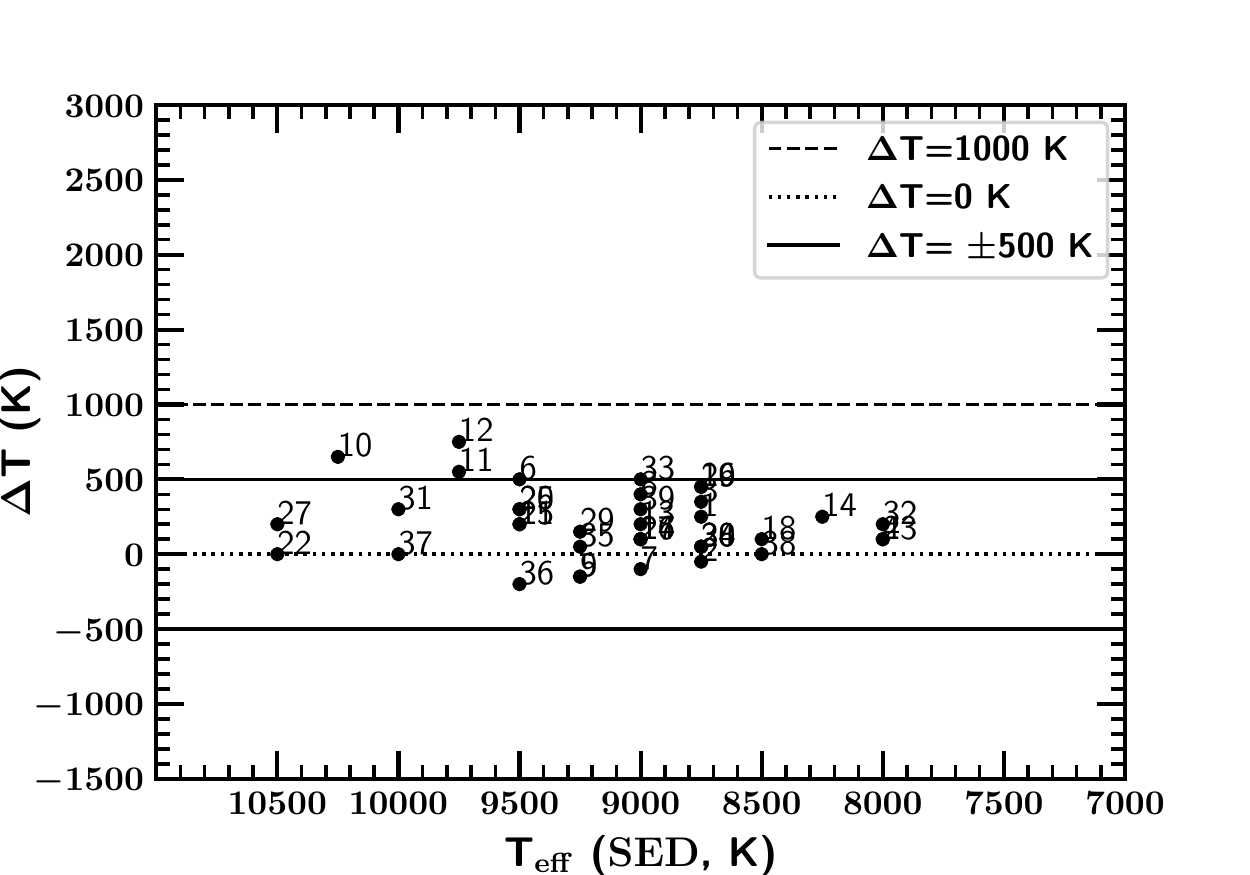} \\ 
        \includegraphics[width=0.498\textwidth]{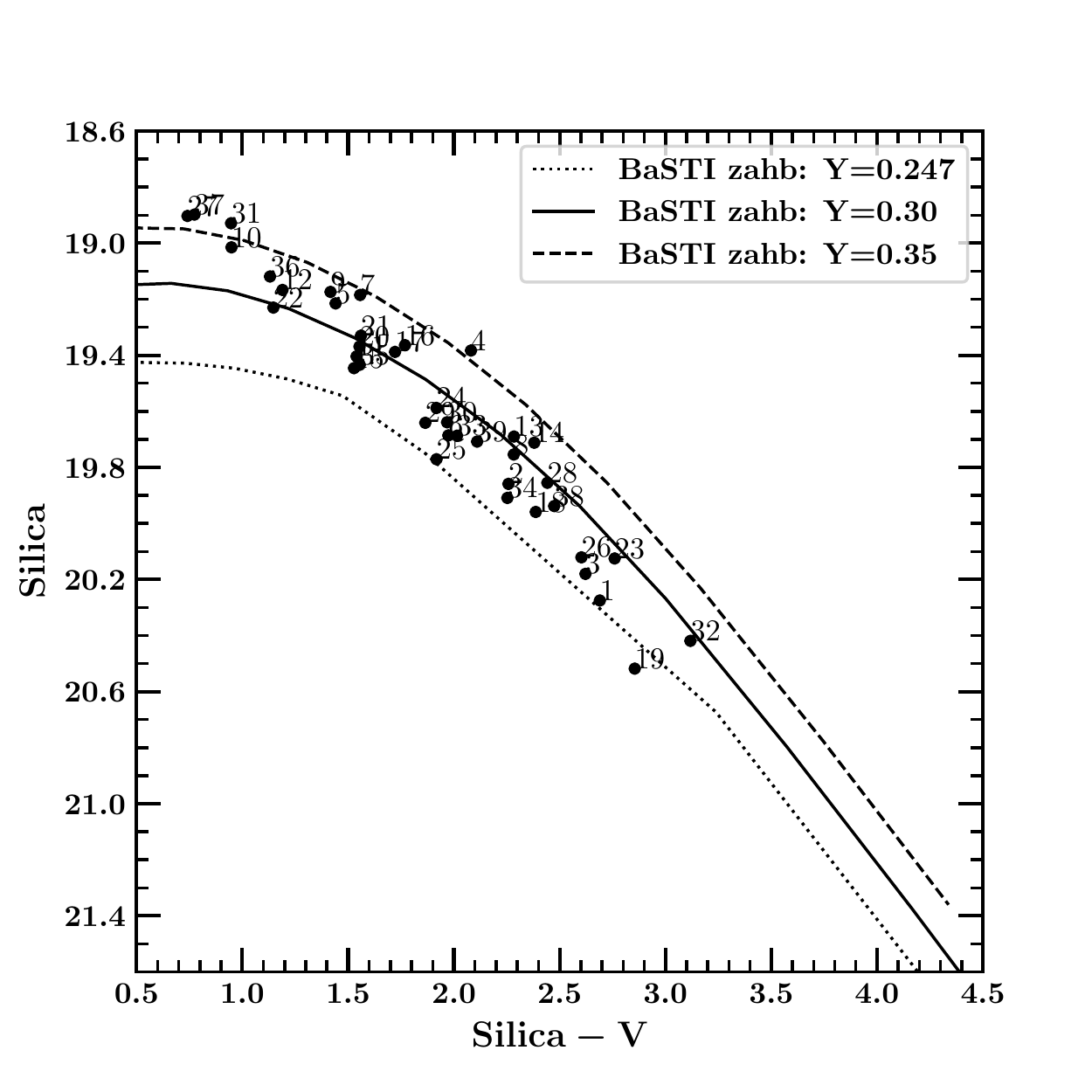}
       \includegraphics[width=0.498\textwidth]{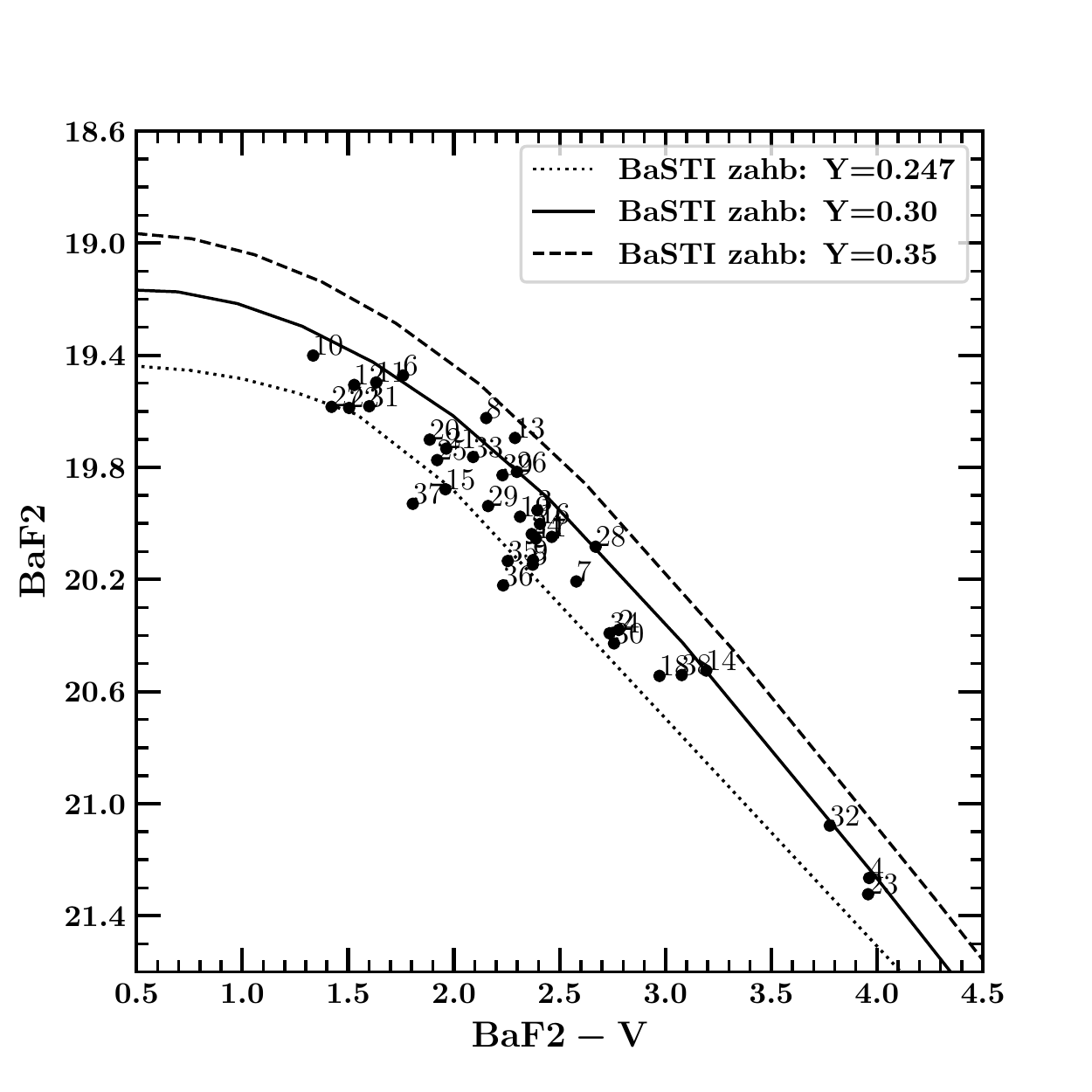}
    \caption{The top left and right panels represent plots of T$_{\mathrm{eff}}^{\mathrm{SED}}$ vs $\Delta$ T (T$_{\mathrm{eff}}^{\mathrm{SED}}$ - T$_{\mathrm{eff}}^{\mathrm{Silica-V}}$) and  T$_{\mathrm{eff}}^{\mathrm{SED}}$ vs $\Delta$ T (T$_{\mathrm{eff}}^{\mathrm{SED}}$ - T$_{\mathrm{eff}}^{\mathrm{BaF2-V}}$), respectively, with each horizontal line showing a particular temperature difference as indicated in the legends.  The black solid circles are the observed BHBs and the reference numbers are the ObjID of BHBs given in  \autoref{tab:BHB_catalog_apendix}. The bottom panels are Silica$-$V vs Silica (left) and BaF2$-$V vs BaF2 (right) CMDs. The BaSTI ZAHB for $Y=0.247,\ 0.300$\ and $0.350$ are shown by dotted, solid, and dashed lines, respectively.}
    \label{fig:TeffDif}
\end{figure*}

\begin{figure*}
    \subfloat[]{\includegraphics[width=0.33\textwidth,]{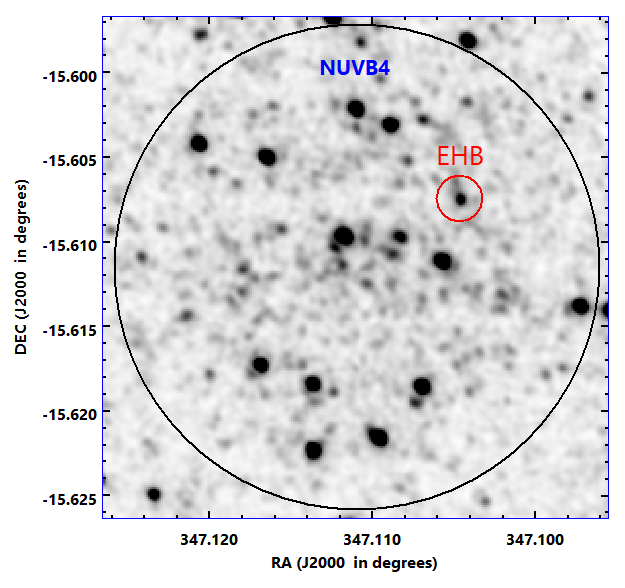}} 
    \subfloat[]{\includegraphics[width=0.33\textwidth]{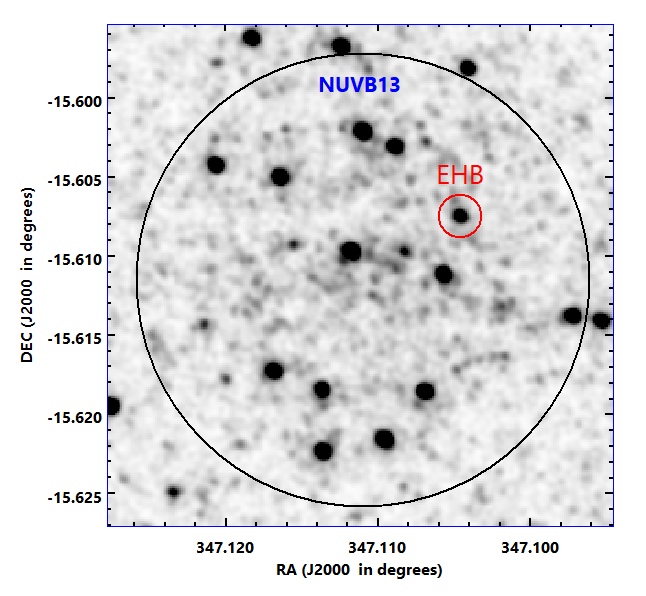}}
    \subfloat[]{\includegraphics[width=0.33\textwidth]{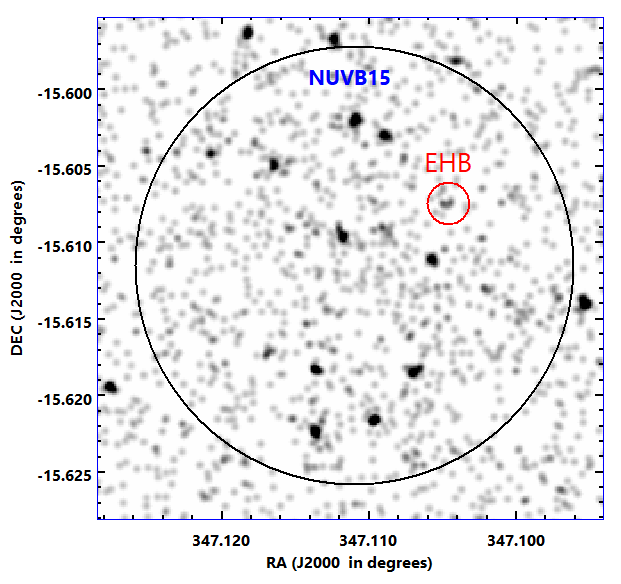}}\\
    \subfloat[]{\includegraphics[width=0.33\textwidth]{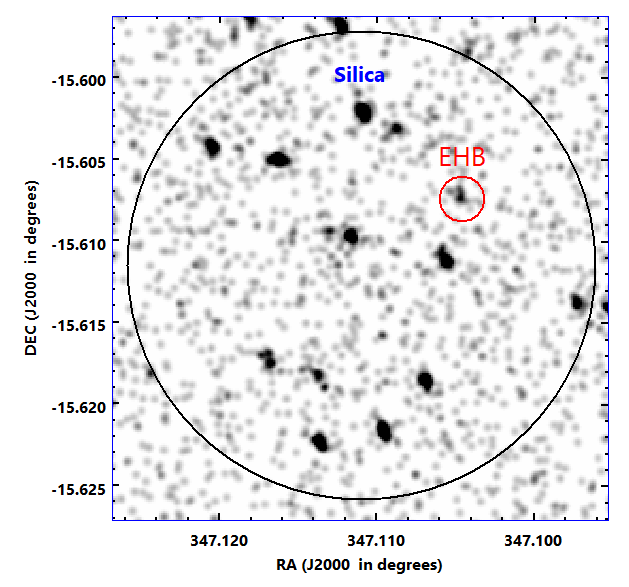}}
    \subfloat[]{\includegraphics[width=0.33\textwidth]{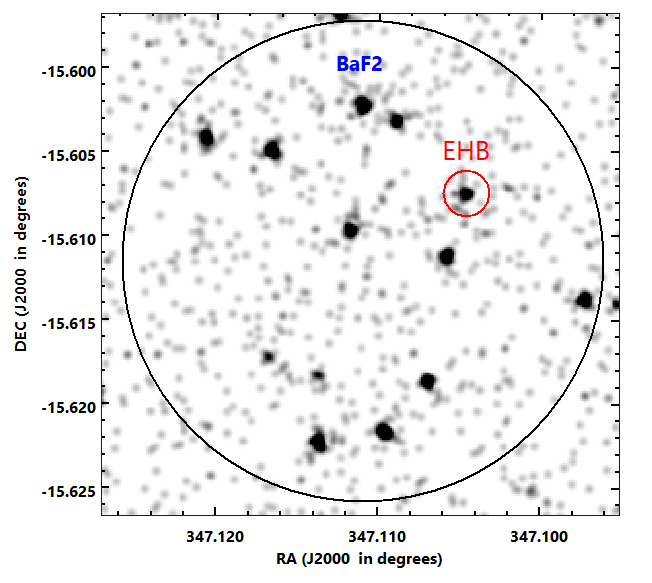}}
    \subfloat[\label{GalFUV}]{\includegraphics[width=0.33\textwidth]{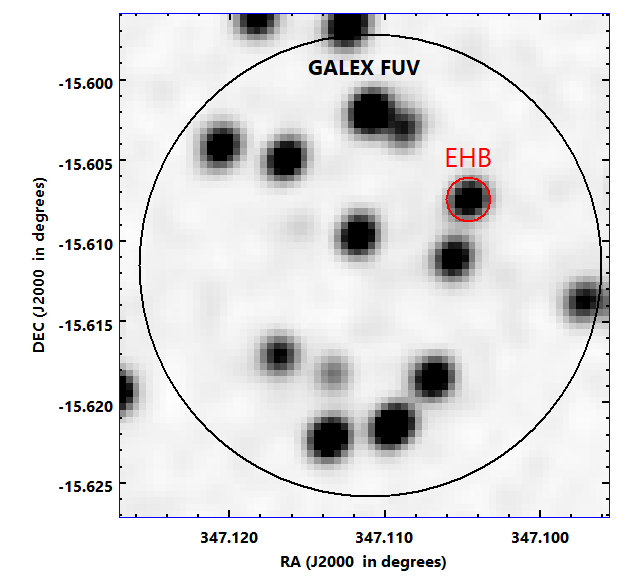}}
    \caption{EHB star observed in different UVIT filters. The filter name is given in blue color on the image and the newly detected EHB star is encircled in a red circle. The black circle shows the core-radius of the cluster. We have also shown the {\em GALEX} FUV observation of the image in Figure (f).}
    \label{fig:ehb}
\end{figure*}

\begin{figure*}
    \centering
    \includegraphics[width=\textwidth]{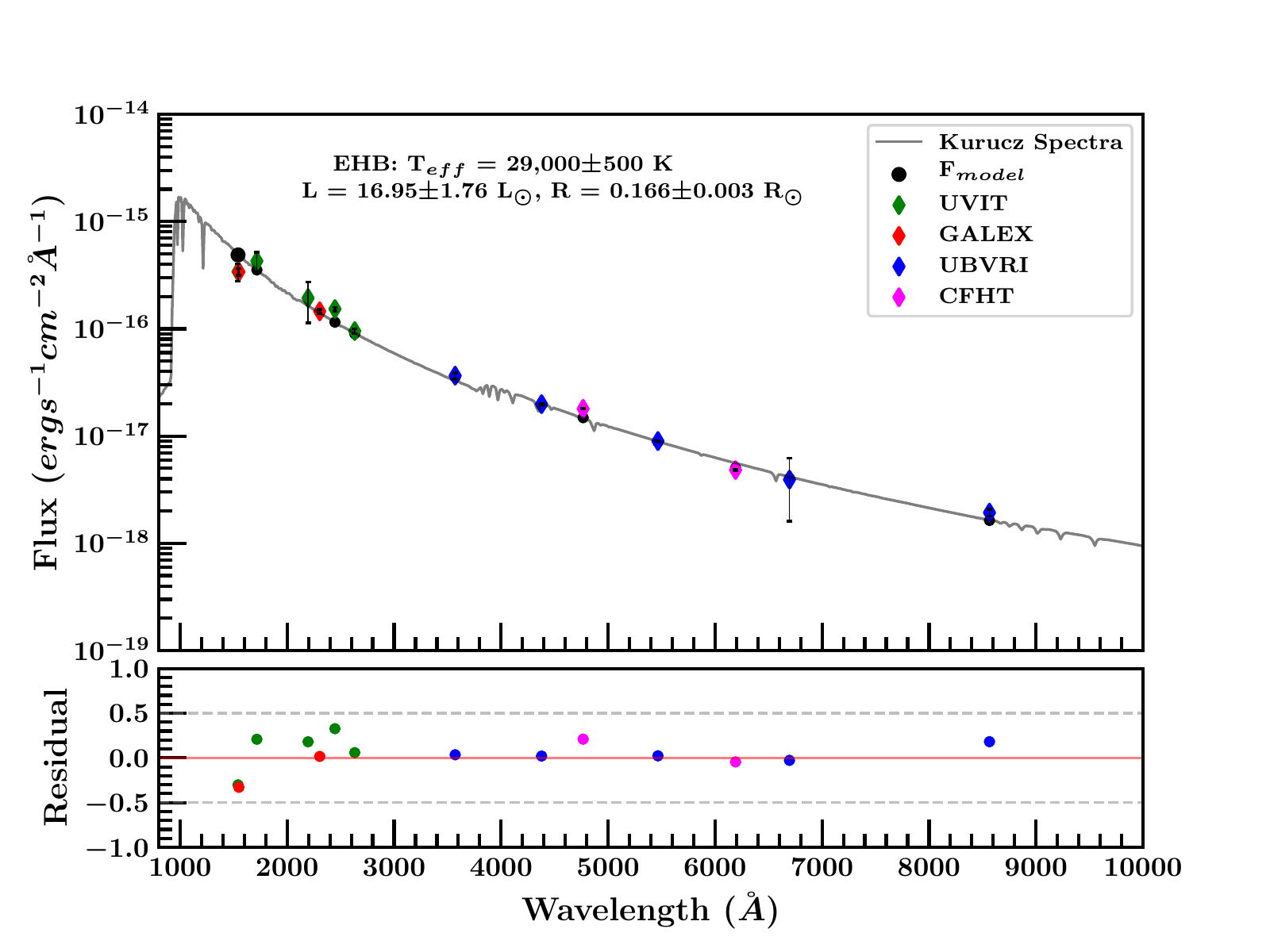}
    \caption{The SED fit of EHB star using the Kurucz model shown in upper panel. The bottom panel shows the residue (deviation) of the observed fluxes from the Kurucz model flux.}
    \label{fig:ehb_sed}
\end{figure*}

\begin{figure}
    \centering
    \includegraphics[width=\columnwidth]{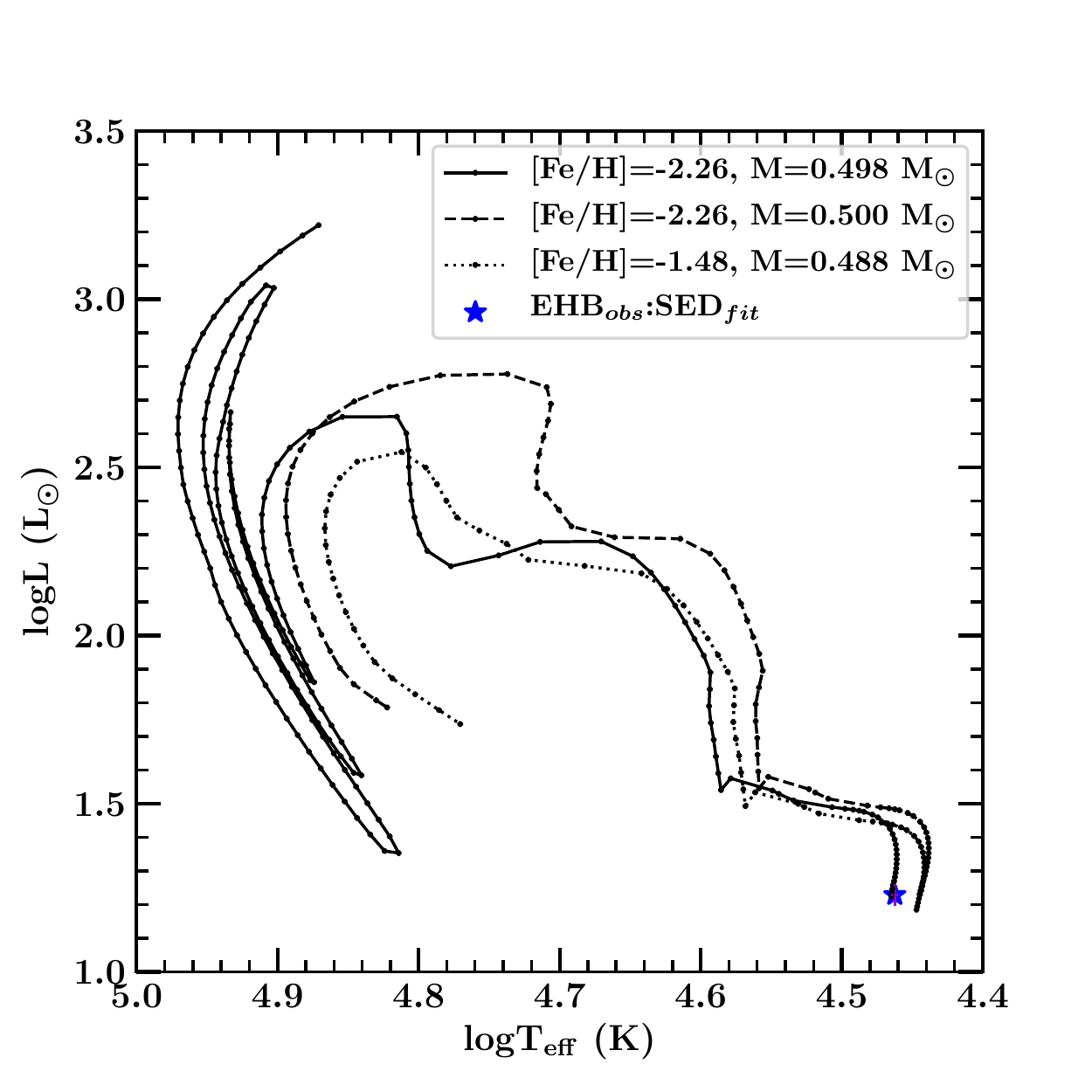}
    \caption{Post-HB evolutionary tracks of the EHB star. The evolutionary tracks with solid and dashed lines are shown for metallicity -2.26 and masses 0.498 M$_\odot$ and 0.500 M$_\odot$, respectively. The dotted line is the evolutionary track of metallicity -1.48 and mass 0.488 M$_\odot$. The Blue asterisk is the position of the newly identified EHB star where $\log$ T$_\mathrm{eff}$ and $\log$ L obtained from SED fit of the EHB.}
    \label{fig:ehb_evol}
\end{figure}

In \autoref{fig:temp_hist}, we have shown the distributions of  T$_{\mathrm{eff}}$ of BHBs derived from both the SEDs fit and color-temperature relations. We see that the BaF2 $-$ GaiaG effective temperature (T$_{\mathrm{eff}}^{\mathrm{BaF2-GaiaG}}$) is closer to the SED effective temperature (T$_{\mathrm{eff}}^{\mathrm{SED}}$) than the Silica $-$ GaiaG effective temperature (T$_{\mathrm{eff}}^{\mathrm{Silica-GaiaG}}$).

We have plotted T$_{\mathrm{eff}}^{\mathrm{SED}}$ $-$ T$_{\mathrm{eff}}^{\mathrm{Silica-V}}$ and T$_{\mathrm{eff}}^{\mathrm{SED}}$ $-$ T$_{\mathrm{eff}}^{\mathrm{BaF2-V}}$ against T$_{\mathrm{eff}}^{\mathrm{SED}}$ with solid circles in the top left and right panels of \autoref{fig:TeffDif}. We have shown the Silica $-$ V vs Silica and BaF2 $-$ V vs BaF2 CMDs, in the  bottom left and right panels, respectively, along with the ZAHB for He-abundances Y = 0.247, 0.300 and 0.350. In the Silica filter (left top and bottom panels), the BHBs which are lying in between isochrones for Y = 0.247 and Y = 0.300 show $\Delta$T of $\pm$500 K. The BHBs observed at He-abundance close to Y = 0.350 have deviation in T$_{\mathrm{eff}}$ of more than 1,000 K from T$_{\mathrm{eff}}^{\mathrm{SED}}$. Some of the BHBs, e.g., HB37, HB27, and HB31 show a deviation of  T$_{\mathrm{eff}}$ more than 1,500 K from the T$_{\mathrm{eff}}^{\mathrm{SED}}$ as they are lying above Y = 0.350 track in Silica$-$V vs Silica CMD. However, with BaF2 filter (right and bottom top panels) we find an overall $\Delta$T of $\pm$500 K and He-abundance of Y = 0.247 to 0.300 for all the observed BHBs .

A difference of 500 K ($\Delta$T) between T$_{\mathrm{eff}}$ obtained from the color-temperature relation of {\em HST} filters and T$_{\mathrm{eff}}$ obtained from spectroscopic observation of BHBs was found by \citet{hb_temp_Lagioia_2015}. A similar comparison of BHBs of NGC 288 was done with $\Delta$T up to 1000 K using UVIT FUV-optical color-temperature relation by \citep{sahu288}. We found a variation of $\Delta$T = $\pm$500 K in between T$_{\mathrm{eff}}^{\mathrm{BaF2-V}}$ (T$_{\mathrm{eff}}^{\mathrm{BaF2-GaiaG}}$) and T$_{\mathrm{eff}}^{\mathrm{SED}}$ of the observed BHBs which is in good agreement with the \citet{hb_temp_Lagioia_2015} and \citet{sahu288} color-temperature estimations. However, we find a deviation of $\Delta$T more than 1000 K between T$_{\mathrm{eff}}^{\mathrm{Silica-V}}$ (T$_{\mathrm{eff}}^{\mathrm{Silica-GaiaG}}$) and T$_{\mathrm{eff}}^{\mathrm{SED}}$ due to the larger sensitivity of Silica magnitudes towards the chemical composition (He, and $\alpha$- enhancements) than the effective temperature of the BHBs. 

\section{Extreme horizontal branch (EHB) star at the center of the cluster}
\label{sec:ehb}

\begin{table*}
    \caption{Various physical parameters derived from the SED fit and evolutionary tracks of the EHB source. The T$_{\mathrm{eff}}$, luminosity and radius are derived from the SED fit and $\log$g, mass of of the source are derived from evolutionary tracks of \citet{Dorman1993}.}
    \label{tab:ehb_sed}
    \centering
    
    \adjustbox{max width=\textwidth} {
    \begin{tabular}{c c c c c c c c c c c}
    \hline
 ID & RA(J2000) & DEC(J2000) & $\chi_{r,sed}^2$ & T$_{\mathrm{eff}}$ & Luminosity  & Radius    & log(g) & M$_{EHB}$      & M$_{core}$ & M$_{envelope}$ \\
  & hh:mm:ss.ss & hh:mm:ss.ss &                    & K         & L$_\odot$   & R$_\odot$ &      & M$_\odot$ & M$_\odot$  & M$_\odot$ \\
    \hline
  EHB01 & 23:08:25.10 & -15:36:26.86 & 12.66 & 29,000 $\pm$ 500 & 16.95 $\pm$ 1.76 & 0.166 $\pm$ 0.003 & 5.72 & 0.498 &  0.495 & 0.003\\ 
    \hline
    \end{tabular} }
\end{table*}

EHBs are visible as faint blue objects in optical filters and occupy the sub-dwarf (sdB/OB) region in the optical CMDs \citep{Heber1987}. However, they are very bright in FUV filters due to their high temperature and represent the hot-flashers in UV CMDs \citep{Dalessandro2011}. The EHBs are formed mainly due to the severe mass-loss of the outer shell of the star in their RGB phase resulting in a very thin H-envelope and a He-core of around 0.5 M$_\odot$. The EHBs evolve towards the white dwarf phases without climbing up towards the AGB and post-AGB phases post its core-helium exhaustion. Consequently, they appear in the region with high temperature and luminosity (UV bright region) which is often known as AGB-manque region \citep{Greggio1990, Dorman1993}.

In the core of the cluster, we found one UV bright source at a distance of 25.48$''$ from the cluster-center. This source was free from any contamination of other cluster members in all the FUV and NUV filters (encircled with a red circle in \autoref{fig:ehb}). When we cross-matched our UVIT sources with the deeper optical photometry from ground-based observations \citep{Stetson2019}, we identified the source as EHB star in all the UV-optical CMDs (blue asterisk in Figures \ref{fig:uvcmd} and \ref{fig:cmd_he}). The UV photometric magnitude of the EHB star ranges from 19.86 mag to 20.54 mag in FUV and NUV filters. Its position and the magnitude in each UV filter are given in \autoref{tab:BHB_catalog_apendix}. This  source also appears very bright in the {\em GALEX} FUV image (Figure \ref{GalFUV}) and is shown in the {\em GALEX} FUV $-$ NUV vs FUV CMD at the position of 0.0 mag in color with FUV magnitude of 20.3 by \citet{Schiavon2012}. However, they did not discuss much about the source in their paper.

We fit the SED of the EHB star using 14 filters from UV to NIR bands to extract T$_{\mathrm{eff}}$, L/L$_\odot$, and radius of the source. Again, we used VOSA SVO SED analyzer to fit the photometric fluxes to the Kurucz model synthetic fluxes for the source considering    metallicity between $-2.0$ and $-1.5$, $\log\, g = 5.0$, $E(B-V) = 0.04$ and distance $d = 24.5 \pm 0.5$ kpc. The SED-fit of the EHB star is shown in the upper panel of \autoref{fig:ehb_sed}. The coloured diamond symbols represent the observed fluxes at the effective wavelengths of the filters whereas the black solids represent the respective model fluxes. The solid line is the best fitted synthetic spectra obtained from the Kurucz model. The lower panel shows the fractional deviation of the observed flux from the model flux and indicated as residual [(= $F_{\mathrm{obs.}} - F_{\mathrm{mod}})/F_{\mathrm{mod}})$]. The T$_{\mathrm{eff}}$ and bolometric luminosity of the EHB star were calculated from the slope of the best fitted spectrum. The radius of the EHB star was calculated from the scaling relation $\left ( \frac{R}{d} \right )^2$ used in the SED fit, where R is the radius of the source and $d$ is the distance. The derived parameters are listed in \autoref{tab:ehb_sed}.    

\citet{Dorman1993} have provided the post-HB evolutionary tracks for $[Fe/H]=-1.48$ and $-2.26$ for metal poor stars. Since the derived T$_{\mathrm{eff}}$ of the observed EHB star is 29,000 K, we see that the post-HB evolutionary track with $[Fe/H]=-2.26$ and M$_{HB}$ of 0.498 M$_\odot$ is lying close to the observed EHB star as seen in \autoref{fig:ehb_evol}. Considering this track as the actual evolutionary track of the observed EHB star, we have also derived its surface gravity, core mass and envelope mass which are given in \autoref{tab:ehb_sed}.

\section{Distribution of sources in the cluster}
\label{sec:discussion}

\begin{figure*}
\centering
    \includegraphics[width=0.42\textwidth, height=0.32\textheight]{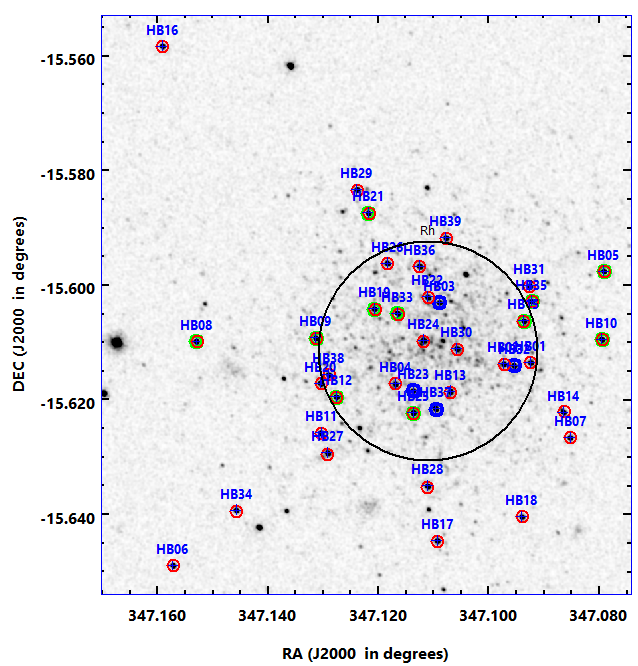}
    \includegraphics[width=0.57\textwidth, height=0.35\textheight]{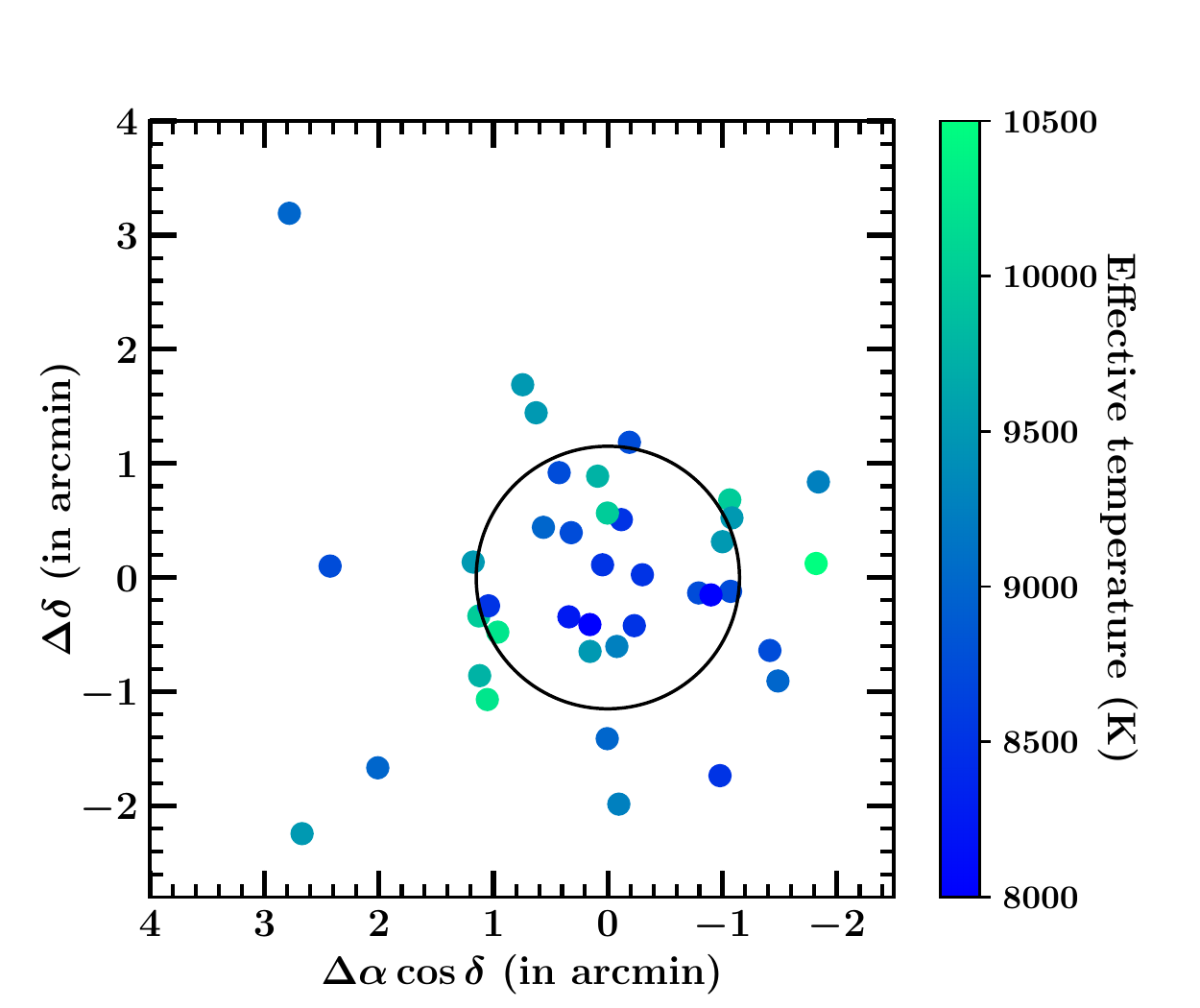}
    \caption{\textit{left panel:} UVIT detections of 39 BHBs are overlaid on the inverted image of NUVB4 filter. We have also over-plotted the {\em GALEX} counterpart BHBs in red circles, the newly identified BHBs in UVIT in blue circles and the 11 available BHBs in SIMBAD  are over-plotted in green circles. The half light radius ($R_h$ = 1.15$'$) is shown in a black circle. \textit{right panel:} The spatial distribution of BHBs at the cluster core on a 2D projection from the cluster center.  X-axis is the distance from the cluster center in RA and Y-axis is the distance from the cluster center in DEC. The black circle denotes the half-light radius of the cluster. The color-bar shows the temperature distribution of the sources.}
    \label{fig:hb_im}
\end{figure*}

\begin{figure}
\centering
    \includegraphics[width=\columnwidth]{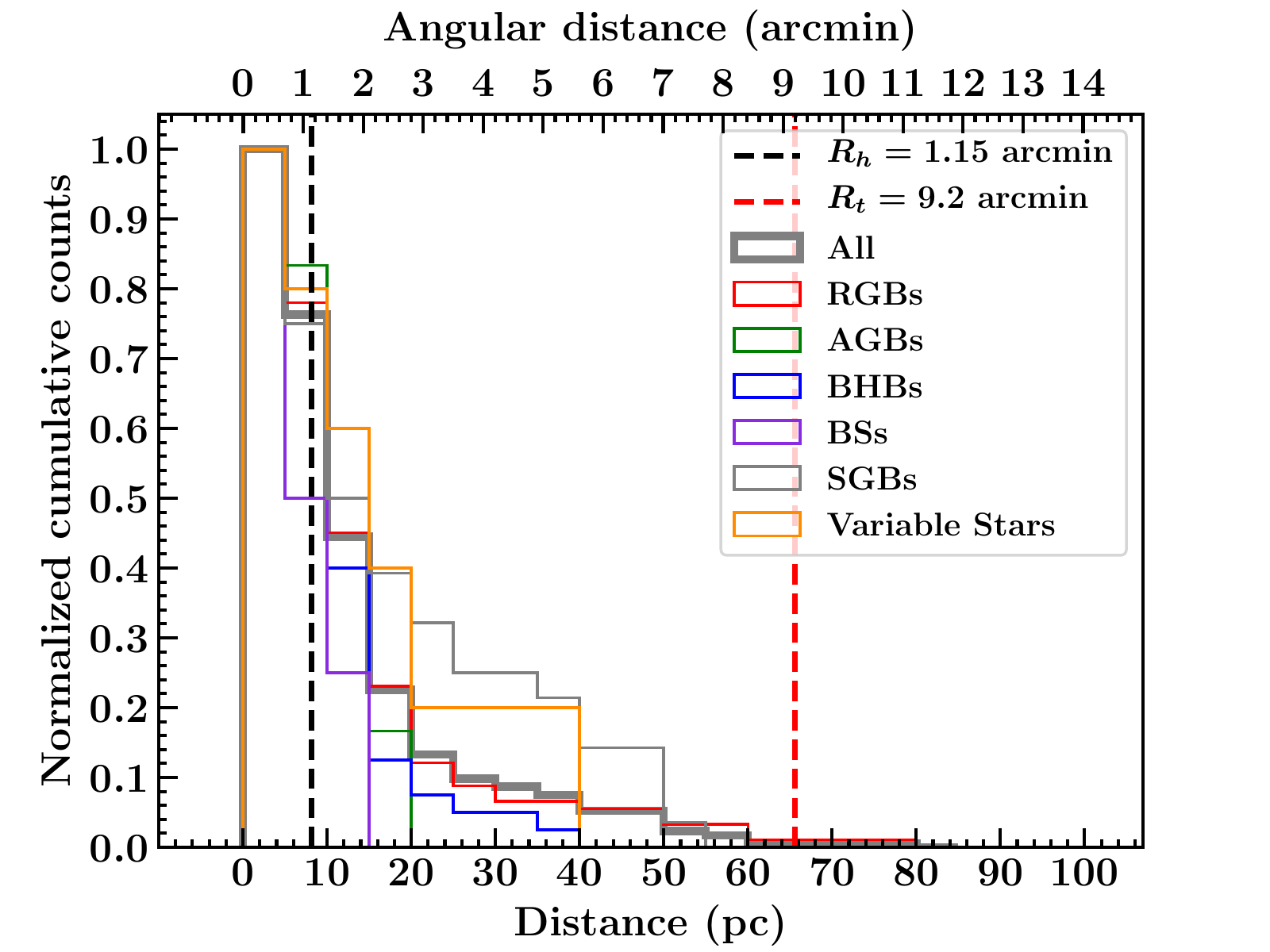}
    \caption{Cumulative radial distribution of all the stellar populations is shown here. The angular distance from the cluster center was converted into the radial distance using a distance-scale of 24.5 kpc. The sources were counted with a bin-size of 5 pc and an inverse normalized cumulative distribution is plotted here. The black dashed line is the half-light radius ($R_h$) and the red dashed line is the tidal-radius ($R_t$) of the cluster. We can see that all the sources (thick gray line) are within the tidal radius of the cluster \citep[$R_t$=9.2$'$,][]{Carballo2012}, 60\% of the BHBs are lying within the half-light radius of the cluster (blue line) and BSs are distributed up to 20 pc from the cluster center.} 
    \label{fig:radial}
\end{figure}

We have classified all the sources of NGC 7492 which have been detected in various filters of UVIT. Interestingly, only the EHB star, BHBs and one RR Lyrae star have been detected in FUV. We have shown the UVIT detected BHBs with the assigned unique IDs (\autoref{tab:BHB_catalog_apendix}) on the inverted image of the NUVB4 filter (left panel, \autoref{fig:hb_im}). We see from  \autoref{fig:hb_im} that 21 of the BHBs are located within the half-light radius of the cluster, and remaining 18 are lying outside the half-light radius. We have over-plotted the {\em GALEX} counterparts of BHBs in red circles and the newly identified BHBs (ObjID: HB03, HB23, HB32 and HB37) in UVIT in blue circles. Only 11 BHBs of our sample are available in SIMBAD \citep[][]{Christlieb2005,Atlee2007} which are denoted by green circles. The BHB, ``HB08'' in the figure was identified as a field member by \cite{Christlieb2005}. However, the proper motion values, $\mu_{RA} = 0.372$ mas/yr and  $\mu_{DEC} = -2.306$ mas/yr, support its candidature as a cluster member in our analysis. All the UVIT detected BHBs are also present in the larger FoV optical band surveys for this cluster. We have shown the temperature distribution of all BHBs in the right panel of \autoref{fig:hb_im}. The T$_{\mathrm{eff}}$ scale is indicated in a vertical color bar. The BHBs have a maximum spread up to 2$'$ from the cluster center but the distribution of T$_{\mathrm{eff}}$ do not show any systematic variation. We have listed the derived temperature and the photometric magnitudes of the UVIT filters of each BHB in \autoref{tab:BHB_catalog_apendix}.

We have shown the radial distribution of all the detected sources in \autoref{fig:radial}. We see that almost 50\% of the sources are within 10 pc (approximately up to the half-light radius of 1.15$'$) of the cluster.  All the BHBs are distributed up to 40 pc from the cluster center. About 80\% of the RGBs are distributed within 20 pc from the center of the cluster and a tail of 5\% RGBs contribute to the outer part of the cluster. \cite{Carballo2012} have shown that this cluster shows a steep density profile and extends the cluster density profile beyond the king's radius of the cluster \cite[$R_t = 8.3'$,][]{Harris2010}. As seen in \autoref{fig:radial}, the presence of RGBs in the outer region of the cluster beyond 8.3$'$ supports the optical observations of \cite{Carballo2012}. 

\begin{figure}
\centering
    \includegraphics[width=\columnwidth]{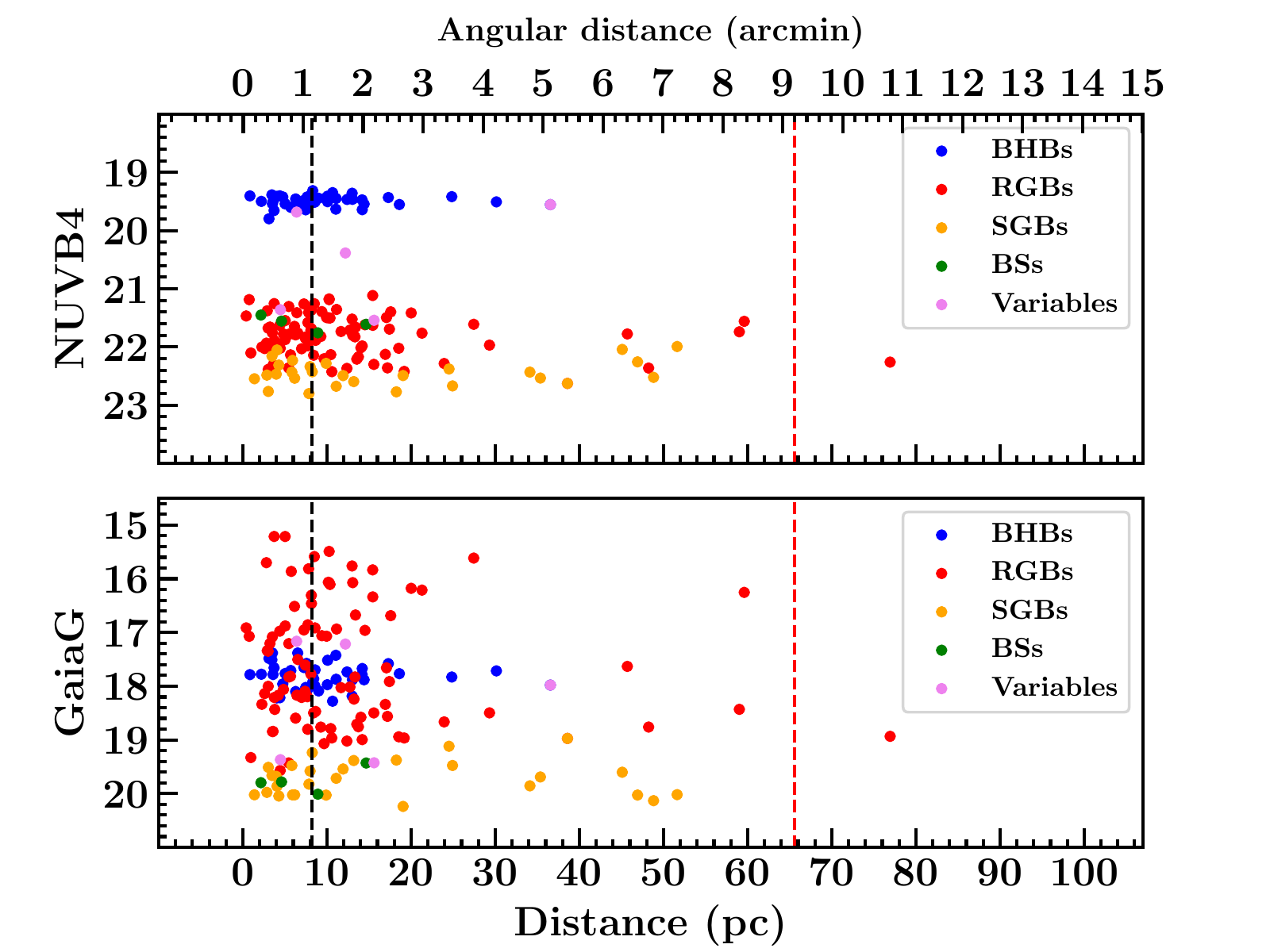}
    \caption{Magnitude distribution over distance from the cluster center of all the stellar populations observed with UVIT filters. The upper panel shows the distribution of sources and their magnitudes observed in the NUVB4 filter and the lower panel shows the distribution of GaiaG magnitudes of their optical counterparts. The black dashed line is the half-light radius ($R_h$) and the red dashed line is the tidal-radius ($R_t$) of the cluster.}   
    \label{fig:mag_dist}
\end{figure}

\autoref{fig:mag_dist} shows the UV (NUVB4 filter) and optical (GaiaG filter) magnitude distributions of the sources over distance from the cluster center. As seen from the figure, BHBs are completely distinguished and there is a clear gap between the brightness of BHBs and RGBs + SGB populations in the UV filter (upper panel, \autoref{fig:mag_dist}). However, in the optical filter (lower panel, \autoref{fig:mag_dist}), all the sources are uniformly distributed at the core of the cluster. This signifies the ability of UV filters to separate out hot sources from the cooler ones. The RR Lyrae stars are lying in the gap between RGBs and BHBs in the UV magnitude distribution. This gap infers the absence of any red-HB (RHB) sources in the cluster \citep[as suggested by][]{Buonanno1987}. The BSs are relatively brighter than the SGBs in UV. Thus, we can infer that there are separate magnitude layers of each evolutionary stages which are hotter than RGBs and SGBs in UV photometry. In the NUV filters (NUVB13 and NUVB4), although we tried to detect the fainter sources up to 23 mag, we could find only those sources which were brighter than MS.

\section{Conclusions}
\label{conclusion}

We have presented a detailed UV photometric analysis for the cluster NGC 7492 with five filters of UVIT. We found 176 sources in NUV and 41 sources in FUV as cluster members and classified them into EHB, BHBs, RGBs, SGBs, variable stars and BSs based upon their positions in UV $-$ optical and optical CMDs. The sources, EHB, BHBs and one RR Lyrae star were detected in multi-band UVIT FUV and NUV filters, whereas sources above MS turn-off were detected only with NUV (NUVB13 and NUVB4) filters. 

The updated BaSTI isochrones and ZAHB, generated using the available cluster parameters, are compared with the observed UV $-$ optical CMDs. The stellar evolution track with metallicity $[Fe/H]=-1.8$, age = 12 Gyrs and a distance modulus of 16.95$\pm$0.05 is fitting well with the observed optical CMDs. The He-abundances of the BHBs of the cluster are estimated using $\alpha$-enhanced ZAHB from BaSTI model for various He-abundances on the UV$-$optical CMDs and also from the R-parameters. The He-abundance value obtained from CMDs is varying from 0.247 to 0.350 whereas R-parameter gives a value of 0.28$\pm$0.05.

We provide a catalog of BHBs along with their effective temperature and UV magnitudes in UVIT filters which will help in improving stellar evolution models and further constrain the structure of the Galaxy. The effective temperatures of all the BHBs were derived using color-temperature relation and SED fitting. The effective temperatures of all the BHBs obtained from both the methods are ranging from 8,000 K to 10,500 K with a deviation of 500 - 1,000 K from each other. Again, the effective temperature derived from the color-temperature relation of BaF2 filter differs a bit from the Silica filter due to the relative shifting of magnitudes between the filters. However, we do not find any systematic pattern in the temperature distribution of the BHBs. 

We have identified one EHB star at the cluster-core at a distance of 25.48$''$ from the center of the cluster. From the  SED fitting we estimated its effective temperature, luminosity and radius to be 29,000$\pm$500 K, 16.95 $\pm$ 1.76  L$_\odot$, and 0.166 $\pm$ 0.003 R$_\odot$, respectively. The post-HB evolutionary tracks from \citet{Dorman1993} are used to estimate surface gravity and mass of the star at ZAHB phase. The $\log$ g of the EHB star is 5.72 whereas its total mass is 0.498 M$_\odot$, with core-mass 0.495 M$_\odot$ and a very thin envelope of mass 0.003 M$_\odot$.

The BHBs in the cluster are distributed uniformly up to 40 pc from the center of the cluster. The uniform distribution of BHBs and their temperature beyond the half-light radius up to 2$'$ from the cluster center supports the dynamics of low density clusters. We find that RGBs are distributed up to the tidal radius of the cluster except one which lies outside the tidal radius suggesting the extension of tidal tails. The hot BHBs are brighter and are clearly distinguished in UV filters whereas they are mixed with the uniform distribution of RGB stars in optical filters.

\section*{Acknowledgements}
We would like to thank the anonymous referee for his/her valuable suggestions and comments. We would also like to thank Prof. Santi Cassisi, INAF - Osservatorio Astronomico d'Abruzzo: Teramo, Abruzzo, Italy, for providing the He-enhanced BaSTI isochrones and ZAHB of UVIT filters early in advance and also for his valuable suggestions and discussions on the He-enhancement of BHBs of the cluster. We would like to thank Dr. Snehalata Sahu of Indian Institute of Astrophysics, Bangalore for fruitful discussions with her on UVIT data reduction. RK would like to acknowledge CSIR Research Fellowship (JRF) Grant No. 09/983(0034)/2019-EMR-1 for the financial support. ACP would like to acknowledge the support by Indian Space Research Organization, Department of Space, Government of India (ISRO RESPOND project  No.ISRO/RES/2/409/17-18). AM thanks DST-INSPIRE (IF150845) for the funding. ACP thanks Inter University centre for Astronomy and Astrophysics (IUCAA), Pune, India for providing facilities to carry out his work. DKO aknowledges the support of the Department of Atomic Energy, Government of India, under Project Identification No. RTI 4002. This publication uses the data from the \mbox{{\em AstroSat}} mission of the Indian Space Research  Organisation (ISRO), archived at the Indian Space Science Data Center (ISSDC). The UVIT data used here was processed by the Payload Operations Centre at IIA. The UVIT is built in collaboration between IIA, IUCAA, TIFR, ISRO and CSA.

\section*{DATA AVAILABILITY}
The data underlying this article will be shared on reasonable request to the corresponding author.

%%%%%%%%%%%%%%%%%%%%%%%%%%%%%%%%%%%%%%%%%%%%%%%%%%

%%%%%%%%%%%%%%%%%%%% REFERENCES %%%%%%%%%%%%%%%%%%

% The best way to enter references is to use BibTeX:
%\bibliographystyle{mnras}
%\bibliography{example} % if your bibtex file is called example.bib
% Alternatively you could enter them by hand, like this:
% This method is tedious and prone to error if you have lots of references

\bibliographystyle{mnras}
\bibliography{7492}

\begin{thebibliography}{}
\makeatletter
\relax
\def\mn@urlcharsother{\let\do\@makeother \do\$\do\&\do\#\do\^\do\_\do\%\do\~}
\def\mn@doi{\begingroup\mn@urlcharsother \@ifnextchar [ {\mn@doi@}
  {\mn@doi@[]}}
\def\mn@doi@[#1]#2{\def\@tempa{#1}\ifx\@tempa\@empty \href
  {http://dx.doi.org/#2} {doi:#2}\else \href {http://dx.doi.org/#2} {#1}\fi
  \endgroup}
\def\mn@eprint#1#2{\mn@eprint@#1:#2::\@nil}
\def\mn@eprint@arXiv#1{\href {http://arxiv.org/abs/#1} {{\tt arXiv:#1}}}
\def\mn@eprint@dblp#1{\href {http://dblp.uni-trier.de/rec/bibtex/#1.xml}
  {dblp:#1}}
\def\mn@eprint@#1:#2:#3:#4\@nil{\def\@tempa {#1}\def\@tempb {#2}\def\@tempc
  {#3}\ifx \@tempc \@empty \let \@tempc \@tempb \let \@tempb \@tempa \fi \ifx
  \@tempb \@empty \def\@tempb {arXiv}\fi \@ifundefined
  {mn@eprint@\@tempb}{\@tempb:\@tempc}{\expandafter \expandafter \csname
  mn@eprint@\@tempb\endcsname \expandafter{\@tempc}}}

\bibitem[\protect\citeauthoryear{{Ambika}, {Parthasarathy}, {Aoki}, {Fujii},
  {Nakada}, {Ita}  \& {Izumiura}}{{Ambika} et~al.}{2004}]{Ambika2004}
{Ambika} S.,  {Parthasarathy} M.,  {Aoki} W.,  {Fujii} T.,  {Nakada} Y.,  {Ita}
  Y.,   {Izumiura} H.,  2004, \mn@doi [\aap] {10.1051/0004-6361:20031759},
  \href {https://ui.adsabs.harvard.edu/abs/2004A&A...417..293A} {417, 293}

\bibitem[\protect\citeauthoryear{{Arakelyan}, {Pilipenko}  \&
  {Libeskind}}{{Arakelyan} et~al.}{2018}]{Arakelyan2018}
{Arakelyan} N.~R.,  {Pilipenko} S.~V.,   {Libeskind} N.~I.,  2018, \mn@doi
  [\mnras] {10.1093/mnras/sty2320}, \href
  {https://ui.adsabs.harvard.edu/abs/2018MNRAS.481..918A} {481, 918}

\bibitem[\protect\citeauthoryear{{Atlee} \& {Gould}}{{Atlee} \&
  {Gould}}{2007}]{Atlee2007}
{Atlee} D.~W.,  {Gould} A.,  2007, \mn@doi [\apj] {10.1086/518467}, \href
  {https://ui.adsabs.harvard.edu/abs/2007ApJ...664...53A} {664, 53}

\bibitem[\protect\citeauthoryear{{Barnes}}{{Barnes}}{1968}]{Barnes1968}
{Barnes} S.~A.,  1968, \mn@doi [\aj] {10.1086/110664}, \href
  {https://ui.adsabs.harvard.edu/abs/1968AJ.....73..579B} {73, 579}

\bibitem[\protect\citeauthoryear{{Baumgardt}, {Hilker}, {Sollima}  \&
  {Bellini}}{{Baumgardt} et~al.}{2019}]{Baumgardt2019}
{Baumgardt} H.,  {Hilker} M.,  {Sollima} A.,   {Bellini} A.,  2019, \mn@doi
  [\mnras] {10.1093/mnras/sty2997}, \href
  {https://ui.adsabs.harvard.edu/abs/2019MNRAS.482.5138B} {482, 5138}

\bibitem[\protect\citeauthoryear{{Bayo}, {Rodrigo}, {Barrado Y Navascu{\'e}s},
  {Solano}, {Guti{\'e}rrez}, {Morales-Calder{\'o}n}  \& {Allard}}{{Bayo}
  et~al.}{2008}]{svo2008}
{Bayo} A.,  {Rodrigo} C.,  {Barrado Y Navascu{\'e}s} D.,  {Solano} E.,
  {Guti{\'e}rrez} R.,  {Morales-Calder{\'o}n} M.,   {Allard} F.,  2008, \mn@doi
  [\aap] {10.1051/0004-6361:200810395}, \href
  {http://adsabs.harvard.edu/abs/2008A%26A...492..277B} {492, 277}

\bibitem[\protect\citeauthoryear{{Behr}}{{Behr}}{2003}]{Behr2003}
{Behr} B.~B.,  2003, \mn@doi [\apjs] {10.1086/377509}, \href
  {https://ui.adsabs.harvard.edu/abs/2003ApJS..149...67B} {149, 67}

\bibitem[\protect\citeauthoryear{Bond \& Alves}{Bond \& Alves}{2001}]{Bond2001}
Bond H.~E.,  Alves D.~R.,  2001, Post-AGB Stars in Globular Clusters and
  Galactic Halos.
Springer Netherlands, Dordrecht, pp 77--82,
  \mn@doi{10.1007/978-94-015-9688-6_11}

\bibitem[\protect\citeauthoryear{{Buonanno}, {Corsi}, {Ferraro}  \& {Fusi
  Pecci}}{{Buonanno} et~al.}{1987}]{Buonanno1987}
{Buonanno} R.,  {Corsi} C.~E.,  {Ferraro} I.,   {Fusi Pecci} F.,  1987, \aaps,
  \href {https://ui.adsabs.harvard.edu/abs/1987A&AS...67..327B} {67, 327}

\bibitem[\protect\citeauthoryear{{Bustos Fierro} \& {Calder{\'o}n}}{{Bustos
  Fierro} \& {Calder{\'o}n}}{2019}]{Fierro2019}
{Bustos Fierro} I.~H.,  {Calder{\'o}n} J.~H.,  2019, \mn@doi [\mnras]
  {10.1093/mnras/stz1879}, \href
  {https://ui.adsabs.harvard.edu/abs/2019MNRAS.488.3024B} {488, 3024}

\bibitem[\protect\citeauthoryear{{Buzzoni}, {Pecci}, {Buonanno}  \&
  {Corsi}}{{Buzzoni} et~al.}{1983}]{Buzzoni1983}
{Buzzoni} A.,  {Pecci} F.~F.,  {Buonanno} R.,   {Corsi} C.~E.,  1983, \aap,
  \href {https://ui.adsabs.harvard.edu/abs/1983A&A...128...94B} {128, 94}

\bibitem[\protect\citeauthoryear{{Carballo-Bello}, {Gieles}, {Sollima},
  {Koposov}, {Mart{\'\i}nez-Delgado}  \& {Pe{\~n}arrubia}}{{Carballo-Bello}
  et~al.}{2012}]{Carballo2012}
{Carballo-Bello} J.~A.,  {Gieles} M.,  {Sollima} A.,  {Koposov} S.,
  {Mart{\'\i}nez-Delgado} D.,   {Pe{\~n}arrubia} J.,  2012, \mn@doi [\mnras]
  {10.1111/j.1365-2966.2011.19663.x}, \href
  {https://ui.adsabs.harvard.edu/abs/2012MNRAS.419...14C} {419, 14}

\bibitem[\protect\citeauthoryear{{Carballo-Bello}, {Sollima},
  {Mart{\'{\i}}nez-Delgado}, {Pila-D{\'{\i}}ez}, {Leaman}, {Fliri}, {Mu{\~n}oz}
   \& {Corral-Santana}}{{Carballo-Bello} et~al.}{2014}]{Carballo2014}
{Carballo-Bello} J.~A.,  {Sollima} A.,  {Mart{\'{\i}}nez-Delgado} D.,
  {Pila-D{\'{\i}}ez} B.,  {Leaman} R.,  {Fliri} J.,  {Mu{\~n}oz} R.~R.,
  {Corral-Santana} J.~M.,  2014, \mn@doi [\mnras] {10.1093/mnras/stu1949},
  \href {https://ui.adsabs.harvard.edu/abs/2014MNRAS.445.2971C} {445, 2971}

\bibitem[\protect\citeauthoryear{{Carballo-Bello} et~al.,}{{Carballo-Bello}
  et~al.}{2018}]{Carballo2018}
{Carballo-Bello} J.~A.,  et~al., 2018, \mn@doi [\mnras]
  {10.1093/mnras/stx3001}, \href
  {https://ui.adsabs.harvard.edu/abs/2018MNRAS.474.4766C} {474, 4766}

\bibitem[\protect\citeauthoryear{{Cardelli}, {Clayton}  \& {Mathis}}{{Cardelli}
  et~al.}{1989}]{cardeli1989}
{Cardelli} J.~A.,  {Clayton} G.~C.,   {Mathis} J.~S.,  1989, \mn@doi [\apj]
  {10.1086/167900}, \href {http://adsabs.harvard.edu/abs/1989ApJ...345..245C}
  {345, 245}

\bibitem[\protect\citeauthoryear{{Castelli} \& {Kurucz}}{{Castelli} \&
  {Kurucz}}{2003}]{kurucz2004}
{Castelli} F.,  {Kurucz} R.~L.,  2003, in {Piskunov} N.,  {Weiss} W.~W.,
  {Gray} D.~F.,  eds,  IAU Symposium Vol. 210, Modelling of Stellar
  Atmospheres. p.~A20 (\mn@eprint {} {astro-ph/0405087})

\bibitem[\protect\citeauthoryear{{Castelli}, {Gratton}  \& {Kurucz}}{{Castelli}
  et~al.}{1997}]{kurucz1997}
{Castelli} F.,  {Gratton} R.~G.,   {Kurucz} R.~L.,  1997, \aap, \href
  {http://adsabs.harvard.edu/abs/1997A%26A...318..841C} {318, 841}

\bibitem[\protect\citeauthoryear{{Chambers} et~al.,}{{Chambers}
  et~al.}{2016}]{Chambers2016}
{Chambers} K.~C.,  et~al., 2016, arXiv e-prints, \href
  {https://ui.adsabs.harvard.edu/abs/2016arXiv161205560C} {p. arXiv:1612.05560}

\bibitem[\protect\citeauthoryear{{Christlieb}, {Beers}, {Thom}, {Wilhelm},
  {Rossi}, {Flynn}, {Wisotzki}  \& {Reimers}}{{Christlieb}
  et~al.}{2005}]{Christlieb2005}
{Christlieb} N.,  {Beers} T.~C.,  {Thom} C.,  {Wilhelm} R.,  {Rossi} S.,
  {Flynn} C.,  {Wisotzki} L.,   {Reimers} D.,  2005, \mn@doi [\aap]
  {10.1051/0004-6361:20041830}, \href
  {https://ui.adsabs.harvard.edu/abs/2005A&A...431..143C} {431, 143}

\bibitem[\protect\citeauthoryear{{Clement}}{{Clement}}{2017}]{Clement2017}
{Clement} C.,  2017, in European Physical Journal Web of Conferences. p. 01021,
  \mn@doi{10.1051/epjconf/201715201021}

\bibitem[\protect\citeauthoryear{{Clement} et~al.,}{{Clement}
  et~al.}{2001}]{Clement2001}
{Clement} C.~M.,  et~al., 2001, \mn@doi [\aj] {10.1086/323719}, \href
  {https://ui.adsabs.harvard.edu/abs/2001AJ....122.2587C} {122, 2587}

\bibitem[\protect\citeauthoryear{{Cohen} \& {Melendez}}{{Cohen} \&
  {Melendez}}{2005}]{Cohen2005}
{Cohen} J.~G.,  {Melendez} J.,  2005, \mn@doi [AJ] {10.1086/427717}, \href
  {http://adsabs.harvard.edu/abs/2005AJ....129.1607C} {129, 1607}

\bibitem[\protect\citeauthoryear{{Cohen}, {Hempel}, {Mauro}, {Geisler},
  {Alonso-Garcia}  \& {Kinemuchi}}{{Cohen} et~al.}{2015}]{Cohen2015}
{Cohen} R.~E.,  {Hempel} M.,  {Mauro} F.,  {Geisler} D.,  {Alonso-Garcia} J.,
  {Kinemuchi} K.,  2015, \mn@doi [\aj] {10.1088/0004-6256/150/6/176}, \href
  {https://ui.adsabs.harvard.edu/abs/2015AJ....150..176C} {150, 176}

\bibitem[\protect\citeauthoryear{{Cote}, {Richer}  \& {Fahlman}}{{Cote}
  et~al.}{1991}]{Cote1991}
{Cote} P.,  {Richer} H.~B.,   {Fahlman} G.~G.,  1991, \mn@doi [AJ]
  {10.1086/115961}, 102, 1358

\bibitem[\protect\citeauthoryear{{Dalessandro}, {Salaris}, {Ferraro},
  {Cassisi}, {Lanzoni}, {Rood}, {Fusi Pecci}  \& {Sabbi}}{{Dalessandro}
  et~al.}{2011}]{Dalessandro2011}
{Dalessandro} E.,  {Salaris} M.,  {Ferraro} F.~R.,  {Cassisi} S.,  {Lanzoni}
  B.,  {Rood} R.~T.,  {Fusi Pecci} F.,   {Sabbi} E.,  2011, \mn@doi [\mnras]
  {10.1111/j.1365-2966.2010.17479.x}, \href
  {https://ui.adsabs.harvard.edu/abs/2011MNRAS.410..694D} {410, 694}

\bibitem[\protect\citeauthoryear{{Dalessandro}, {Schiavon}, {Rood}, {Ferraro},
  {Sohn}, {Lanzoni}  \& {O'Connell}}{{Dalessandro}
  et~al.}{2012}]{Dalessandro2012}
{Dalessandro} E.,  {Schiavon} R.~P.,  {Rood} R.~T.,  {Ferraro} F.~R.,  {Sohn}
  S.~T.,  {Lanzoni} B.,   {O'Connell} R.~W.,  2012, \mn@doi [AJ]
  {10.1088/0004-6256/144/5/126}, 144, 126

\bibitem[\protect\citeauthoryear{{Dalessandro}, {Salaris}, {Ferraro},
  {Mucciarelli}  \& {Cassisi}}{{Dalessandro} et~al.}{2013}]{Dalessandro2013}
{Dalessandro} E.,  {Salaris} M.,  {Ferraro} F.~R.,  {Mucciarelli} A.,
  {Cassisi} S.,  2013, \mn@doi [\mnras] {10.1093/mnras/sts644}, \href
  {https://ui.adsabs.harvard.edu/abs/2013MNRAS.430..459D} {430, 459}

\bibitem[\protect\citeauthoryear{{Dorman}, {Rood}  \& {O'Connell}}{{Dorman}
  et~al.}{1993}]{Dorman1993}
{Dorman} B.,  {Rood} R.~T.,   {O'Connell} R.~W.,  1993, \mn@doi [\apj]
  {10.1086/173511}, \href
  {https://ui.adsabs.harvard.edu/abs/1993ApJ...419..596D} {419, 596}

\bibitem[\protect\citeauthoryear{{Ferraro}}{{Ferraro}}{2003}]{Ferraro2003}
{Ferraro} F.~R.,  2003, Memorie della Societa Astronomica Italiana Supplementi,
  \href {https://ui.adsabs.harvard.edu/abs/2003MSAIS...3...80F} {3, 80}

\bibitem[\protect\citeauthoryear{{Figuera Jaimes}, {Arellano Ferro}, {Bramich},
  {Giridhar}  \& {Kuppuswamy}}{{Figuera Jaimes} et~al.}{2013}]{Figuera2013}
{Figuera Jaimes} R.,  {Arellano Ferro} A.,  {Bramich} D.~M.,  {Giridhar} S.,
  {Kuppuswamy} K.,  2013, \mn@doi [AAP] {10.1051/0004-6361/201220824}, 556, A20

\bibitem[\protect\citeauthoryear{{Forbes} \& {Bridges}}{{Forbes} \&
  {Bridges}}{2010}]{Forbes2010}
{Forbes} D.~A.,  {Bridges} T.,  2010, \mn@doi [\mnras]
  {10.1111/j.1365-2966.2010.16373.x}, \href
  {https://ui.adsabs.harvard.edu/abs/2010MNRAS.404.1203F} {404, 1203}

\bibitem[\protect\citeauthoryear{{Freeman} \& {Norris}}{{Freeman} \&
  {Norris}}{1981}]{Freeman1981}
{Freeman} K.~C.,  {Norris} J.,  1981, \mn@doi [\araa]
  {10.1146/annurev.aa.19.090181.001535}, \href
  {https://ui.adsabs.harvard.edu/abs/1981ARA&A..19..319F} {19, 319}

\bibitem[\protect\citeauthoryear{{Gaia Collaboration} et~al.,}{{Gaia
  Collaboration} et~al.}{2018a}]{GaiaCatalog2018}
{Gaia Collaboration} et~al., 2018a, \mn@doi [\aap]
  {10.1051/0004-6361/201833051}, \href
  {https://ui.adsabs.harvard.edu/abs/2018A&A...616A...1G} {616, A1}

\bibitem[\protect\citeauthoryear{{Gaia Collaboration} et~al.,}{{Gaia
  Collaboration} et~al.}{2018b}]{gaia_kinematics}
{Gaia Collaboration} et~al., 2018b, \mn@doi [\aap]
  {10.1051/0004-6361/201832698}, 616, A12

\bibitem[\protect\citeauthoryear{{Green}, {Schlafly}, {Zucker}, {Speagle}  \&
  {Finkbeiner}}{{Green} et~al.}{2019}]{Green2019}
{Green} G.~M.,  {Schlafly} E.,  {Zucker} C.,  {Speagle} J.~S.,   {Finkbeiner}
  D.,  2019, \mn@doi [\apj] {10.3847/1538-4357/ab5362}, \href
  {https://ui.adsabs.harvard.edu/abs/2019ApJ...887...93G} {887, 93}

\bibitem[\protect\citeauthoryear{{Greggio} \& {Renzini}}{{Greggio} \&
  {Renzini}}{1990}]{Greggio1990}
{Greggio} L.,  {Renzini} A.,  1990, \mn@doi [\apj] {10.1086/169384}, \href
  {https://ui.adsabs.harvard.edu/abs/1990ApJ...364...35G} {364, 35}

\bibitem[\protect\citeauthoryear{{Harris}}{{Harris}}{2010}]{Harris2010}
{Harris} W.~E.,  2010, arXiv e-prints, \href
  {http://adsabs.harvard.edu/abs/2010arXiv1012.3224H} {}

\bibitem[\protect\citeauthoryear{{Harris} \& {Racine}}{{Harris} \&
  {Racine}}{1979}]{Harris1979}
{Harris} W.~E.,  {Racine} R.,  1979, \mn@doi [\araa]
  {10.1146/annurev.aa.17.090179.001325}, \href
  {https://ui.adsabs.harvard.edu/abs/1979ARA&A..17..241H} {17, 241}

\bibitem[\protect\citeauthoryear{{Harris}, {Nemec}  \& {Hesser}}{{Harris}
  et~al.}{1983}]{Harris1983}
{Harris} H.~C.,  {Nemec} J.~M.,   {Hesser} J.~E.,  1983, \mn@doi [\pasp]
  {10.1086/131153}, \href
  {https://ui.adsabs.harvard.edu/abs/1983PASP...95..256H} {95, 256}

\bibitem[\protect\citeauthoryear{{Heber}}{{Heber}}{1987}]{Heber1987}
{Heber} U.,  1987, in {Philip} A.~G.~D.,  {Hayes} D.~S.,   {Liebert} J.~W.,
  eds, IAU Colloq. 95: Second Conference on Faint Blue Stars. pp 79--88

\bibitem[\protect\citeauthoryear{{Hidalgo} et~al.,}{{Hidalgo}
  et~al.}{2018}]{Hidalgo2018}
{Hidalgo} S.~L.,  et~al., 2018, \mn@doi [\apj] {10.3847/1538-4357/aab158},
  \href {https://ui.adsabs.harvard.edu/abs/2018ApJ...856..125H} {856, 125}

\bibitem[\protect\citeauthoryear{{Jain}, {Vig}  \& {Ghosh}}{{Jain}
  et~al.}{2019}]{Jain2019}
{Jain} R.,  {Vig} S.,   {Ghosh} S.~K.,  2019, \mn@doi [\mnras]
  {10.1093/mnras/stz544}, \href
  {https://ui.adsabs.harvard.edu/abs/2019MNRAS.485.2877J} {485, 2877}

\bibitem[\protect\citeauthoryear{Jasniewicz \& Parthasarathy}{Jasniewicz \&
  Parthasarathy}{2009}]{Jasniewicz2009a}
Jasniewicz G.,  Parthasarathy M.,  2009, in Richtler T.,  Larsen S.,  eds,
  Globular Clusters - Guides to Galaxies. Springer Berlin Heidelberg, Berlin,
  Heidelberg, pp 35--36, \mn@doi{10.1007/978-3-540-76961-3_10}

\bibitem[\protect\citeauthoryear{{Jasniewicz}, {de Laverny}, {Parthasarathy},
  {L{\`e}bre}  \& {Th{\'e}venin}}{{Jasniewicz} et~al.}{2004}]{Jasniewicz2004}
{Jasniewicz} G.,  {de Laverny} P.,  {Parthasarathy} M.,  {L{\`e}bre} A.,
  {Th{\'e}venin} F.,  2004, \mn@doi [\aap] {10.1051/0004-6361:20034504}, \href
  {https://ui.adsabs.harvard.edu/abs/2004A&A...423..353J} {423, 353}

\bibitem[\protect\citeauthoryear{{Keller}, {Mackey}  \& {Da Costa}}{{Keller}
  et~al.}{2012}]{Keller2012}
{Keller} S.~C.,  {Mackey} D.,   {Da Costa} G.~S.,  2012, \mn@doi [\apj]
  {10.1088/0004-637X/744/1/57}, \href
  {https://ui.adsabs.harvard.edu/abs/2012ApJ...744...57K} {744, 57}

\bibitem[\protect\citeauthoryear{{Kruijssen}, {Pfeffer}, {Reina-Campos},
  {Crain}  \& {Bastian}}{{Kruijssen} et~al.}{2019}]{Kruijssen2019}
{Kruijssen} J.~M.~D.,  {Pfeffer} J.~L.,  {Reina-Campos} M.,  {Crain} R.~A.,
  {Bastian} N.,  2019, \mn@doi [\mnras] {10.1093/mnras/sty1609}, \href
  {https://ui.adsabs.harvard.edu/abs/2019MNRAS.486.3180K} {486, 3180}

\bibitem[\protect\citeauthoryear{{Kumar}, {Ghosh}  \& et al.}{{Kumar}
  et~al.}{2012a}]{Kumar2012}
{Kumar} A.,  {Ghosh} S.~K.,   et al. 2012a, in Space Telescopes and
  Instrumentation 2012: Ultraviolet to Gamma Ray. p. 84431N (\mn@eprint {arXiv}
  {1208.4670}), \mn@doi{10.1117/12.924507}

\bibitem[\protect\citeauthoryear{{Kumar}, {Ghosh}  \& et al.}{{Kumar}
  et~al.}{2012b}]{Kumar2012a}
{Kumar} A.,  {Ghosh} S.~K.,   et al. 2012b, in Space Telescopes and
  Instrumentation 2012: Ultraviolet to Gamma Ray. p. 84434R (\mn@eprint {}
  {1208.4672}), \mn@doi{10.1117/12.924147}

\bibitem[\protect\citeauthoryear{{Kumar}, {Pradhan}, {Parthasarathy}, {Ojha},
  {Mohapatra}, {Murthy}  \& {Cassisi}}{{Kumar} et~al.}{2020a}]{Kumar2020a}
{Kumar} R.,  {Pradhan} A.~C.,  {Parthasarathy} M.,  {Ojha} D.~K.,  {Mohapatra}
  A.,  {Murthy} J.,   {Cassisi} S.,  2020a, arXiv e-prints, \href
  {https://ui.adsabs.harvard.edu/abs/2020arXiv201207318K} {p. arXiv:2012.07318}

\bibitem[\protect\citeauthoryear{{Kumar}, {Pradhan}, {Parthasarathy}, {Ojha},
  {Mohapatra}  \& {Murthy}}{{Kumar} et~al.}{2020b}]{Kumar2020}
{Kumar} R.,  {Pradhan} A.~C.,  {Parthasarathy} M.,  {Ojha} D.~K.,  {Mohapatra}
  A.,   {Murthy} J.,  2020b, in {Bragaglia} A.,  {Davies} M.,  {Sills} A.,
  {Vesperini} E.,  eds,  IAU Symposium Vol. 351, Star Clusters: From the Milky
  Way to the Early Universe. pp 464--467 (\mn@eprint {arXiv} {1908.02512}),
  \mn@doi{10.1017/S1743921319007373}

\bibitem[\protect\citeauthoryear{Lagioia, Dalessandro, Ferraro, Salaris,
  Lanzoni, Pietrinferni  \& Cassisi}{Lagioia
  et~al.}{2015}]{hb_temp_Lagioia_2015}
Lagioia E.~P.,  Dalessandro E.,  Ferraro F.~R.,  Salaris M.,  Lanzoni B.,
  Pietrinferni A.,   Cassisi S.,  2015, \mn@doi [The Astrophysical Journal]
  {10.1088/0004-637x/800/1/52}, 800, 52

\bibitem[\protect\citeauthoryear{{Lee}, {Lee}, {Fahlman}  \& {Sung}}{{Lee}
  et~al.}{2004}]{Lee2004}
{Lee} K.~H.,  {Lee} H.~M.,  {Fahlman} G.~G.,   {Sung} H.,  2004, \mn@doi [AJ]
  {10.1086/425547}, 128, 2838

\bibitem[\protect\citeauthoryear{{Marino} et~al.,}{{Marino}
  et~al.}{2014}]{Marino2014}
{Marino} A.~F.,  et~al., 2014, \mn@doi [\mnras] {10.1093/mnras/stt1993}, \href
  {https://ui.adsabs.harvard.edu/abs/2014MNRAS.437.1609M} {437, 1609}

\bibitem[\protect\citeauthoryear{{Massari}, {Koppelman}  \& {Helmi}}{{Massari}
  et~al.}{2019}]{Massari2019}
{Massari} D.,  {Koppelman} H.~H.,   {Helmi} A.,  2019, \mn@doi [\aap]
  {10.1051/0004-6361/201936135}, \href
  {https://ui.adsabs.harvard.edu/abs/2019A&A...630L...4M} {630, L4}

\bibitem[\protect\citeauthoryear{{Moehler}}{{Moehler}}{2010}]{Moehler2010}
{Moehler} S.,  2010, \memsai, \href
  {https://ui.adsabs.harvard.edu/abs/2010MmSAI..81..838M} {81, 838}

\bibitem[\protect\citeauthoryear{{Moehler}, {Landsman}  \&
  {Napiwotzki}}{{Moehler} et~al.}{1998}]{Moehler1998}
{Moehler} S.,  {Landsman} W.,   {Napiwotzki} R.,  1998, \aap, \href
  {https://ui.adsabs.harvard.edu/abs/1998A&A...335..510M} {335, 510}

\bibitem[\protect\citeauthoryear{{Moehler}, {Landsman}, {Lanz}  \& {Miller
  Bertolami}}{{Moehler} et~al.}{2019}]{Moehler2019}
{Moehler} S.,  {Landsman} W.~B.,  {Lanz} T.,   {Miller Bertolami} M.~M.,  2019,
  \mn@doi [\aap] {10.1051/0004-6361/201935694}, \href
  {https://ui.adsabs.harvard.edu/abs/2019A&A...627A..34M} {627, A34}

\bibitem[\protect\citeauthoryear{{Mu{\~n}oz}, {C{\^o}t{\'e}}, {Santana},
  {Geha}, {Simon}, {Oyarz{\'u}n}, {Stetson}  \& {Djorgovski}}{{Mu{\~n}oz}
  et~al.}{2018}]{Munoz2018}
{Mu{\~n}oz} R.~R.,  {C{\^o}t{\'e}} P.,  {Santana} F.~A.,  {Geha} M.,  {Simon}
  J.~D.,  {Oyarz{\'u}n} G.~A.,  {Stetson} P.~B.,   {Djorgovski} S.~G.,  2018,
  \mn@doi [\apj] {10.3847/1538-4357/aac168}, \href
  {https://ui.adsabs.harvard.edu/abs/2018ApJ...860...65M} {860, 65}

\bibitem[\protect\citeauthoryear{{Navarrete}, {Belokurov}  \&
  {Koposov}}{{Navarrete} et~al.}{2017}]{Navarrete2017}
{Navarrete} C.,  {Belokurov} V.,   {Koposov} S.~E.,  2017, \mn@doi [ApJL]
  {10.3847/2041-8213/aa72e1}, 841, L23

\bibitem[\protect\citeauthoryear{{Parthasarathy}, {Jasniewicz}, {Aoki}  \&
  {Takeda}}{{Parthasarathy} et~al.}{2012}]{Parthasarathy2012}
{Parthasarathy} M.,  {Jasniewicz} G.,  {Aoki} W.,   {Takeda} Y.,  2012, in
  {Aoki} W.,  {Ishigaki} M.,  {Suda} T.,  {Tsujimoto} T.,   {Arimoto} N.,  eds,
   Astronomical Society of the Pacific Conference Series Vol. 458, Galactic
  Archaeology: Near-Field Cosmology and the Formation of the Milky Way. p.~237

\bibitem[\protect\citeauthoryear{{Piotto} et~al.,}{{Piotto}
  et~al.}{2015}]{Piotto2015}
{Piotto} G.,  et~al., 2015, \mn@doi [\aj] {10.1088/0004-6256/149/3/91}, \href
  {https://ui.adsabs.harvard.edu/abs/2015AJ....149...91P} {149, 91}

\bibitem[\protect\citeauthoryear{{Posti} \& {Helmi}}{{Posti} \&
  {Helmi}}{2019}]{Posti2019}
{Posti} L.,  {Helmi} A.,  2019, \mn@doi [\aap] {10.1051/0004-6361/201833355},
  \href {https://ui.adsabs.harvard.edu/abs/2019A&A...621A..56P} {621, A56}

\bibitem[\protect\citeauthoryear{Postma \& Leahy}{Postma \&
  Leahy}{2017}]{Postma2017}
Postma J.~E.,  Leahy D.,  2017, \mn@doi [\pasp] {10.1088/1538-3873/aa8800},
  129, 115002

\bibitem[\protect\citeauthoryear{{Rahna}, {Murthy}, {Safonova}, {Sutaria},
  {Gudennavar}  \& {Bubbly}}{{Rahna} et~al.}{2017}]{Rahna2017}
{Rahna} P.~T.,  {Murthy} J.,  {Safonova} M.,  {Sutaria} F.,  {Gudennavar}
  S.~B.,   {Bubbly} S.~G.,  2017, \mn@doi [\mnras] {10.1093/mnras/stx1748},
  \href {https://ui.adsabs.harvard.edu/abs/2017MNRAS.471.3028R} {471, 3028}

\bibitem[\protect\citeauthoryear{{Riffel}, {Ruschel-Dutra}, {Pastoriza},
  {Rodr{\'\i}guez-Ardila}, {Santos}, {Bonatto}  \& {Ducati}}{{Riffel}
  et~al.}{2011}]{Riffel2011}
{Riffel} R.,  {Ruschel-Dutra} D.,  {Pastoriza} M.~G.,  {Rodr{\'\i}guez-Ardila}
  A.,  {Santos} J.~F.~C. J.,  {Bonatto} C.~J.,   {Ducati} J.~R.,  2011, \mn@doi
  [\mnras] {10.1111/j.1365-2966.2010.17647.x}, \href
  {https://ui.adsabs.harvard.edu/abs/2011MNRAS.410.2714R} {410, 2714}

\bibitem[\protect\citeauthoryear{{Rosenberg}}{{Rosenberg}}{2000}]{Rosenberg2000}
{Rosenberg} A.,  2000, \mn@doi [\pasp] {10.1086/316556}, \href
  {https://ui.adsabs.harvard.edu/abs/2000PASP..112..575R} {112, 575}

\bibitem[\protect\citeauthoryear{{Sahu}, {Subramaniam}, {C{\^o}t{\'e}}, {Rao}
  \& {Stetson}}{{Sahu} et~al.}{2019}]{sahu288}
{Sahu} S.,  {Subramaniam} A.,  {C{\^o}t{\'e}} P.,  {Rao} N.~K.,   {Stetson}
  P.~B.,  2019, \mn@doi [\mnras] {10.1093/mnras/sty2679}, \href
  {http://adsabs.harvard.edu/abs/2019MNRAS.482.1080S} {482, 1080}

\bibitem[\protect\citeauthoryear{{Sarajedini} et~al.,}{{Sarajedini}
  et~al.}{2007}]{Sarajedini2007}
{Sarajedini} A.,  et~al., 2007, \mn@doi [\aj] {10.1086/511979}, \href
  {https://ui.adsabs.harvard.edu/abs/2007AJ....133.1658S} {133, 1658}

\bibitem[\protect\citeauthoryear{{Schiavon} et~al.,}{{Schiavon}
  et~al.}{2012}]{Schiavon2012}
{Schiavon} R.~P.,  et~al., 2012, \mn@doi [AJ] {10.1088/0004-6256/143/5/121},
  143, 121

\bibitem[\protect\citeauthoryear{Shapley}{Shapley}{1918}]{Shapley1918}
Shapley H.,  1918, \mn@doi [Publications of the Astronomical Society of the
  Pacific] {10.1086/122686}, 30, 42

\bibitem[\protect\citeauthoryear{{Shapley}}{{Shapley}}{1920}]{Shapley1920}
{Shapley} H.,  1920, \mn@doi [\apj] {10.1086/142560}, \href
  {https://ui.adsabs.harvard.edu/abs/1920ApJ....52...73S} {52}

\bibitem[\protect\citeauthoryear{{Shapley} \& {Sawyer}}{{Shapley} \&
  {Sawyer}}{1927}]{Shapley1927}
{Shapley} H.,  {Sawyer} H.~B.,  1927, Harvard College Observatory Bulletin,
  \href {https://ui.adsabs.harvard.edu/abs/1927BHarO.849...11S} {849, 11}

\bibitem[\protect\citeauthoryear{{Stetson}, {Pancino}, {Zocchi}, {Sanna}  \&
  {Monelli}}{{Stetson} et~al.}{2019}]{Stetson2019}
{Stetson} P.~B.,  {Pancino} E.,  {Zocchi} A.,  {Sanna} N.,   {Monelli} M.,
  2019, \mn@doi [\mnras] {10.1093/mnras/stz585}, \href
  {https://ui.adsabs.harvard.edu/abs/2019MNRAS.485.3042S} {485, 3042}

\bibitem[\protect\citeauthoryear{{Subramaniam}, {Tandon}  \& et.
  al.}{{Subramaniam} et~al.}{2016}]{Subramaniam2016}
{Subramaniam} A.,  {Tandon} S.~N.,   et. al. 2016, in Space Telescopes and
  Instrumentation 2016: Ultraviolet to Gamma Ray. p. 99051F (\mn@eprint {}
  {1608.01073}), \mn@doi{10.1117/12.2235271}

\bibitem[\protect\citeauthoryear{{Subramaniam} et~al.,}{{Subramaniam}
  et~al.}{2017}]{Subramaniam2017}
{Subramaniam} A.,  et~al., 2017, \mn@doi [\aj] {10.3847/1538-3881/aa94c3},
  \href {https://ui.adsabs.harvard.edu/abs/2017AJ....154..233S} {154, 233}

\bibitem[\protect\citeauthoryear{{Tandon}, {Subramaniam}, {Girish}  \& et.
  al.}{{Tandon} et~al.}{2017}]{Tandon2017}
{Tandon} S.~N.,  {Subramaniam} A.,  {Girish} V.,   et. al. 2017, \mn@doi [\aj]
  {10.3847/1538-3881/aa8451}, 154, 128

\bibitem[\protect\citeauthoryear{{Tandon} et~al.,}{{Tandon}
  et~al.}{2020}]{Tandon2020}
{Tandon} S.~N.,  et~al., 2020, \mn@doi [\aj] {10.3847/1538-3881/ab72a3}, \href
  {https://ui.adsabs.harvard.edu/abs/2020AJ....159..158T} {159, 158}

\bibitem[\protect\citeauthoryear{{Tenorio-Tagle}, {Mu{\~n}oz-Tu{\~n}{\'o}n},
  {Cassisi}  \& {Silich}}{{Tenorio-Tagle} et~al.}{2016}]{Tenorio2016}
{Tenorio-Tagle} G.,  {Mu{\~n}oz-Tu{\~n}{\'o}n} C.,  {Cassisi} S.,   {Silich}
  S.,  2016, \mn@doi [\apj] {10.3847/0004-637X/825/2/118}, \href
  {https://ui.adsabs.harvard.edu/abs/2016ApJ...825..118T} {825, 118}

\bibitem[\protect\citeauthoryear{{Tenorio-Tagle}, {Silich}, {Palou{\v{s}}},
  {Mu{\~n}oz-Tu{\~n}{\'o}n}  \& {W{\"u}nsch}}{{Tenorio-Tagle}
  et~al.}{2019}]{Tenorio2019}
{Tenorio-Tagle} G.,  {Silich} S.,  {Palou{\v{s}}} J.,
  {Mu{\~n}oz-Tu{\~n}{\'o}n} C.,   {W{\"u}nsch} R.,  2019, \mn@doi [\apj]
  {10.3847/1538-4357/ab2455}, \href
  {https://ui.adsabs.harvard.edu/abs/2019ApJ...879...58T} {879, 58}

\bibitem[\protect\citeauthoryear{{Vanderbeke} et~al.,}{{Vanderbeke}
  et~al.}{2014}]{Vanderbeke2014}
{Vanderbeke} J.,  et~al., 2014, \mn@doi [\mnras] {10.1093/mnras/stt2002}, \href
  {https://ui.adsabs.harvard.edu/abs/2014MNRAS.437.1725V} {437, 1725}

\bibitem[\protect\citeauthoryear{{Vasiliev}}{{Vasiliev}}{2019}]{Vasiliev2018}
{Vasiliev} E.,  2019, \mn@doi [\mnras] {10.1093/mnras/stz171}, \href
  {https://ui.adsabs.harvard.edu/abs/2019MNRAS.484.2832V} {484, 2832}

\bibitem[\protect\citeauthoryear{{Villanova}, {Geisler}, {Piotto}  \&
  {Gratton}}{{Villanova} et~al.}{2012}]{Villanova2012}
{Villanova} S.,  {Geisler} D.,  {Piotto} G.,   {Gratton} R.~G.,  2012, \mn@doi
  [\apj] {10.1088/0004-637X/748/1/62}, \href
  {https://ui.adsabs.harvard.edu/abs/2012ApJ...748...62V} {748, 62}

\bibitem[\protect\citeauthoryear{{Zinn}, {Newell}  \& {Gibson}}{{Zinn}
  et~al.}{1972}]{Zinn1972}
{Zinn} R.~J.,  {Newell} E.~B.,   {Gibson} J.~B.,  1972, \aap, \href
  {https://ui.adsabs.harvard.edu/abs/1972A&A....18..390Z} {18, 390}

\makeatother
\end{thebibliography}

%%%%%%%%%%%%%%%%%%%%%%%%%%%%%%%%%%%%%%%%%%%%%%%%%%

%%%%%%%%%%%%%%%%% APPENDICES %%%%%%%%%%%%%%%%%%%%%
\appendix
\label{apendix}

% Don't change these lines
\bsp	% typesetting comment
\label{lastpage}
\end{document}